\documentclass[sigplan,10pt,screen]{acmart}

\renewcommand\footnotetextcopyrightpermission[1]{} 
\setcopyright{none}
\usepackage{hyperref}
\usepackage{tikz}
\usepackage{color}
\usepackage{xcolor}
\definecolor{strawberry}{rgb}{1.0, 0.26, 0.64}
\definecolor{ruby}{rgb}{0.88, 0.07, 0.37}
\definecolor{princetonorange}{rgb}{1.0, 0.56, 0.0}

\usepackage{enumitem}
\usepackage{epsfig,endnotes}
\usepackage[export]{adjustbox}
\usepackage{tabularx}
\usepackage{appendix}

\usepackage{lineno}
\usepackage[linesnumbered,ruled,noline,noend]{algorithm2e}
\usepackage{graphicx}
\usepackage{float}
\usepackage{xspace}
\usepackage{multirow}
\usepackage{comment}
\usepackage{fancybox, fancyvrb, calc}
\usepackage{subcaption}
\usepackage{amsmath}
\usepackage{amsthm}

\usepackage{amssymb}
\usepackage{mathtools}
\usepackage{natbib}
\usepackage{pbox}
\usepackage{rotating}
\usepackage{hhline}

\usepackage{microtype}
\usepackage{epstopdf}
\usepackage[normalem]{ulem}

\newcommand{\cut}[1]{}

\newcommand{\paraspace}{\vspace{0.05in}}
\newcommand{\parab}[1]{\paraspace\noindent{\bf #1} }

\newcommand{\eg}{e.g., }

\newcommand{\ie}{i.e., }

\usepackage{pifont}
\newcommand{\xmark}{\ding{55}}%
\newcommand{\yes}{{\checkmark}}
\newcommand{\no}{\xmark}

\newcommand{\sys}{\textsc{Astraea}\xspace}

\begin{document}

\title{Towards Fair and Efficient Learning-based Congestion Control}

\thanks{$^*$Equal contribution.}

\author{\Large
Xudong Liao$^*$, Han Tian$^*$, Chaoliang Zeng, Xinchen Wan, Kai Chen
}
\affiliation{
  \institution{iSING Lab, Hong Kong University of Science and Technology\country{}}
}

\renewcommand{\shortauthors}{Xudong Liao et al.}
\renewcommand{\authors}{Xudong Liao, Han Tian, Chaoliang Zeng, Xinchen Wan, Kai Chen}

\begin{abstract}
Recent years have witnessed a plethora of learning-based solutions for congestion control (CC) that demonstrate better performance over traditional TCP schemes. However, they fail to provide consistently good convergence properties, including {\em fairness}, {\em fast convergence} and {\em stability}, due to the mismatch between their objective functions and these properties. Despite being intuitive, integrating these properties into existing learning-based CC is challenging, because: 1) their training environments are designed for the performance optimization of single flow but incapable of cooperative multi-flow optimization, and 2) there is no directly measurable metric to represent these properties into the training objective function.

We present \sys, a new learning-based congestion control that ensures fast convergence to fairness with stability. At the heart of \sys is a multi-agent deep reinforcement learning framework that explicitly optimizes these convergence properties during the training process by enabling the learning of interactive policy between multiple competing flows, while maintaining high performance. We further build a faithful multi-flow environment that emulates the competing behaviors of concurrent flows, explicitly expressing convergence properties to enable their optimization during training. We have fully implemented \sys and our comprehensive experiments show that \sys can quickly converge to fairness point and exhibit better stability than its counterparts. For example, \sys achieves near-optimal bandwidth sharing (i.e., fairness) when multiple flows compete for the same bottleneck, delivers up to 8.4$\times$ faster convergence speed and 2.8$\times$ smaller throughput deviation, while achieving comparable or even better performance over prior solutions. 

\end{abstract}

  


\maketitle
\pagestyle{plain}

\section{Introduction}
Internet congestion control (CC) remains an active field of research in both academia and industry. An expected end-to-end Internet CC algorithm should preserve two abilities. First, the algorithm should achieve high performance, \ie high throughput, low latency, and few congestion loss in various network conditions. Second, it needs to provide good convergence properties, \ie rate allocation in a fair and efficient manner\footnote{The efficient manner means all participant flows should fully utilize the bottleneck capacity.}, and fast convergence with stability. Despite three decades efforts, it is still hard to achieve the above requirements simultaneously. 

Classical TCP CC algorithms~\cite{reno, cubic, bbr, vegas, compound, fast, xcp} has been notorious for performance degradation when their assumptions about congestion and packet-level events are violated~\cite{allegro, exp-cc}. To address this problem, a recent evolved thread of research has provided us with a plethora of clean-slate learning-based CC approaches~\cite{allegro, vivace, aurora, orca, owl, mocc, spine, genet}. These learning-based schemes achieve performance goals by defining an objective function consisting of throughput, latency, and loss rate. Offline training~\cite{aurora, orca} or online optimization~\cite{vivace} are then applied to guide the rate control. Learning-based approaches enable the potential to adapt CC to various network conditions and hence achieve consistently high performance.

While substantially improving over traditional TCPs in terms of performance, these learning-based CC schemes have so far shown little improvement on convergence properties. For example, as shown in \S\ref{sec:motiv}, Vivace~\cite{vivace}, an online learning-based scheme, shows a hard tradeoff between convergence speed and stability, although it is designed with these metrics in mind. Moreover, Aurora~\cite{aurora}, a reinforcement learning (RL)-based scheme, is fairness-agnostic and shares no bandwidth with competing flows. Orca~\cite{orca} identifies the similar convergence issue, and proposes a coupling idea by incorporating classical TCP into RL-based control. Therefore, it can preserve certain fairness provided by TCP. However, based on our evaluations (\S\ref{sec:exp:conv}), we find that Orca achieves instable convergence. The reason behind is that Orca's RL part may hurt the theoretical guarantee of fairness from AIMD by suppressing the loss event.


The crux is that existing learning-based CC schemes \emph{do not directly} optimize for the convergence properties, because their decentralized learning paradigms only optimize local performance objectives based on local observations. They try to reach the convergence indirectly by optimizing for the performance goals, which often turns out to be sub-optimal. For instance, Vivace empirically maximizes a latency-sensitive objective and guarantees a fair equilibrium when competing flows follow the same optimization method. It, however, may converge slowly and lead to an inefficiently fair allocation in the wild Internet, which was uncovered by \cite{orca}. Moreover, it is difficult to tune the control knobs to achieve a good tradeoff between performance goals and various convergence characteristics, as shown in \S\ref{sec:motiv}, because these knobs are generally not related to the convergence properties semantically. 

Therefore, we ask: {\em is it possible to automatically develop a CC algorithm that can quickly converge to fair rate allocation with stability, while maintaining high throughput, low latency, and low packet loss rate?}

In this paper, we answer this question affirmatively with \sys.
\sys is an end-to-end RL-based CC scheme for Internet that \emph{explicitly} introduces convergence metrics including fairness, convergence speed, and stability, in addition to throughput, latency, and loss, into the optimization objective.
Despite being intuitive, \sys calls for a fundamental change in the training framework design compared with prior learning-based CC systems: \sys adopts a centralized and cooperative multi-agent training paradigm considering the dynamics between multiple flows to optimize global performance.


In \sys{}, each flow is guided by an RL \emph{agent}, which perceives packet statistics as the input of its control policy, and enforces back the new \emph{\mbox{cwnd}}. During the training, \sys collects observations from all concurrent participants in the network, and issues a reward signal encoding the performance and convergence metrics to the RL policy for reinforcement. Eventually, through joint optimization on these goals, the RL agent learns a control policy that works collaboratively with competing flows to achieve a fair, efficient and stable equilibrium without using any pre-defined control rules or hardwired modeling about the network.

Realizing such multi-agent DRL in CC, however, is challenging. First, before \sys, there is no faithful training playground for CC that supports directly studying multiple competing flows. In order to enable explicit optimizations on the aforementioned metrics, we need to feed the training algorithm with the global information of all participant flows. We address this challenge by designing a novel multi-flow environment (\S\ref{sec:design:env}). It can establish multiple concurrent flows in the network and provide measurements on convergence properties. Our environment supports flexible flow configurations to enable complex network conditions, simulating the dynamics of the wild Internet and performing as an effective training suite for \sys algorithm.

Second, it is not easy to encode convergence properties into RL reward objective, as it lacks directly measurable metrics (like throughput and latency) to evaluate these properties. To solve this problem, we design novel representations of convergence properties, \ie fairness and stability, with readily achievable average throughput of each flow (\S\ref{sec:design:agent}). As a result, we can simply represent the global reward function in a linear combination of metrics that we focus on, \ie performance requirements and convergence properties.

Third, we find that training \sys with classical DRL algorithms suffers from large variance, as the complexity of the environment increases with the number of concurrent flows in the bottleneck. To tackle this challenge, we leverage a customized multi-agent DRL training algorithm (\S\ref{sec:design:training}), which incorporates observed empirical trajectories of all active flows and calculates the global reward as a signal. Compared with the standard RL training procedure, it introduces global information involving all active flows to reduce the estimation variance of the value-action function, which will stabilize and accelerate the multi-agent RL training.

We have implemented \sys and corresponding environment, and performed efficient distributed training for speedup. We integrate the trained \sys algorithm with Linux kernel TCP (\S\ref{sec:impl}). Extensive experiments (\S\ref{sec:exp}) over emulation and real-world Internet demonstrate that \sys significantly improves the convergence properties while preserving high performance in a diverse range of network conditions and multiple bottleneck scenarios. For example, in the experiments of multiple homogeneous flows, \sys shows near-optimal fairness, achieves the average Jain index of 0.991 and improves the convergence speed and stability by up to 8.4 $\times$ and 3.3$\times$ compared with existing TCP variants and learning-based schemes. In addition, \sys delivers comparable RTT fairness and TCP friendliness with existing learning-based schemes. In real-world experiments, \sys always defines the frontiers of high throughput and low latency: it delivers 3.1$\times$ higher throughput than Orca and 1.4 $\times$ lower latency than BBR.


In summary, we make the following contributions.
\begin{itemize}[leftmargin=*]
	\item We present \sys, a multi-agent DRL solution for CC that aims to directly optimize for fairness, convergence speed and stability as well as the performance goals. \sys is a showcase of solving distributed problems with centralized learning methods.
	\item We design and implement a multi-flow environment for training \sys. To the best of our knowledge, this is the first CC playground that implements state and action passing mechanism, and global information gathering for the multi-agent training algorithm. The environment will also facilitate the further study on convergence behaviors of heterogeneous CC schemes.
	\item We evaluate \sys over real-world and emulated environment and show that \sys significantly improves convergence properties while preserving high performance.
\end{itemize}

The code of \sys{} is available at \url{https://github.com/HKUST-SING/astraea}.

\section{Motivation}\label{sec:motiv}
Driven by the tremendous successes achieved by machine learning in computer systems and networks~\cite{auto,pensieve, decima}, recent proposals seek to leverage learning-based methods to design efficient CC algorithms~\cite{vivace, aurora, orca}. While these schemes have demonstrated superior performance over traditional CC algorithms by providing high throughput, low latency, and low loss rate, they fall short of achieving well-behaved convergence properties, \ie \emph{fast convergence to fairness with stability and efficiency}, as summarized in Table~\ref{table:cc-comp}. All of them cannot fulfill fairness, stability and responsiveness simultaneously.
\begin{table}[t!]
    \begin{tabular}{c c c c}
    \toprule
    Algorithm & Fairness & Fast Convergence & Stability \\
    \hline
    Aurora~\cite{aurora} & \no  & \yes  & \no  \\
    Vivace~\cite{vivace} & \yes & \no   & \no  \\
    Orca~\cite{orca}  & \yes & \yes  & \no  \\
    \sys   & \yes & \yes  & \yes \\
    \bottomrule
    \end{tabular}
    \caption{Comparison of learning-based algorithms. None of the existing learning-based algorithms can satisfy all three requirements. 
    Note that Orca is coupled with CUBIC by default, which results in the issue of stability. Besides, as Orca's RL module does not consider optimizing for fairness, there is no effort on this metric beyond TCP.
    }
    \label{table:cc-comp}
\end{table}

The reason of the failure, we argue, is that their predefined local utility functions are often hard to align with these desiderata, especially \emph{global fairness}. 
In the following, we first give a brief introduction of two representative learning-based congestion control: Aurora~\cite{aurora} and Vivace~\cite{vivace}, and exemplify the objective discrepancy by experimental results.

Aurora is among the first to introduce deep reinforcement learning (DRL) to solve congestion control problem, whose core is a carefully designed reward function (Equation~\ref{eq:aurora-reward}) that encodes the network utility. It performs offline training to learn a control policy that maximizes the accumulated reward and uses this policy to direct the sending process. On the other side, Vivace defines a latency-aware utility function (Equation~\ref{eq:vivace-utility}, $x_{i}$ and $L_{i}$ the sending rate and loss rate in monitor interval $i$, respectively.) and controls sending rate by performing gradient-ascent online learning to optimize this utility function. While achieving high performance, they both have different issues related to convergence properties.

\begin{align}
	r &= 10 \times \mbox{throughput} - 1000 \times \mbox{latency} - 2000 \times \mbox{loss} \label{eq:aurora-reward}\\
	u &= x_{i}^{0.9} - 900 \times x_{i}\frac{d(RTT_{i})}{dT} - 11.25 \times x_{i} \times L_{i}\label{eq:vivace-utility}
\end{align}

\begin{figure}[t!]
    \centering
    \begin{subfigure}[b]{0.49\linewidth}
        \centering
        \includegraphics[width=\linewidth]{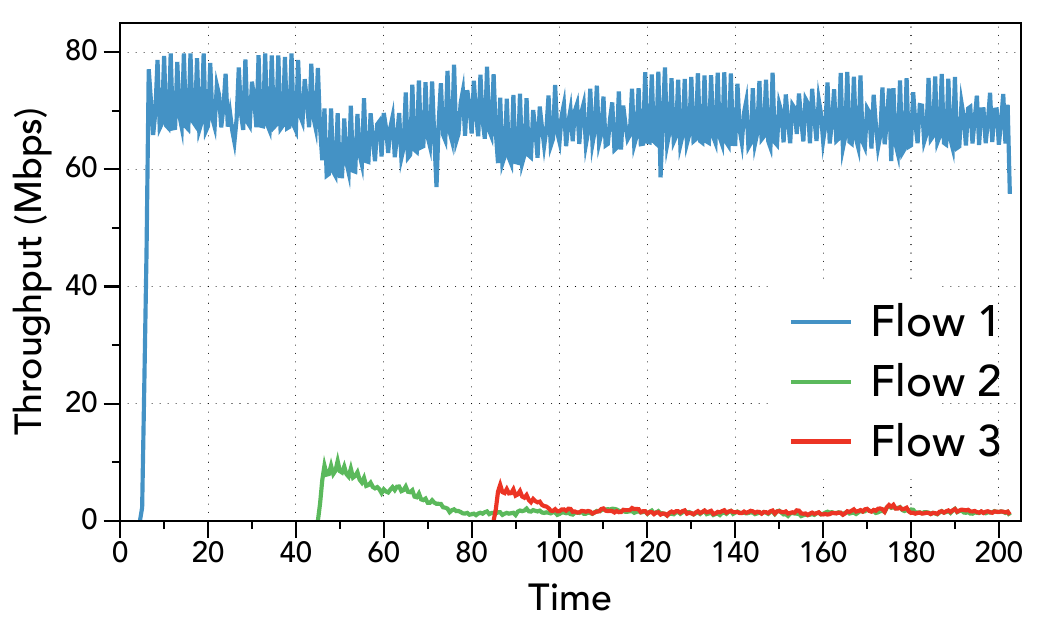}
        \caption{Aurora is very unfair.}
        \label{fig:motiv:aurora-unfair}
    \end{subfigure}
    \begin{subfigure}[b]{0.49\linewidth}
        \centering
        \includegraphics[width=\linewidth]{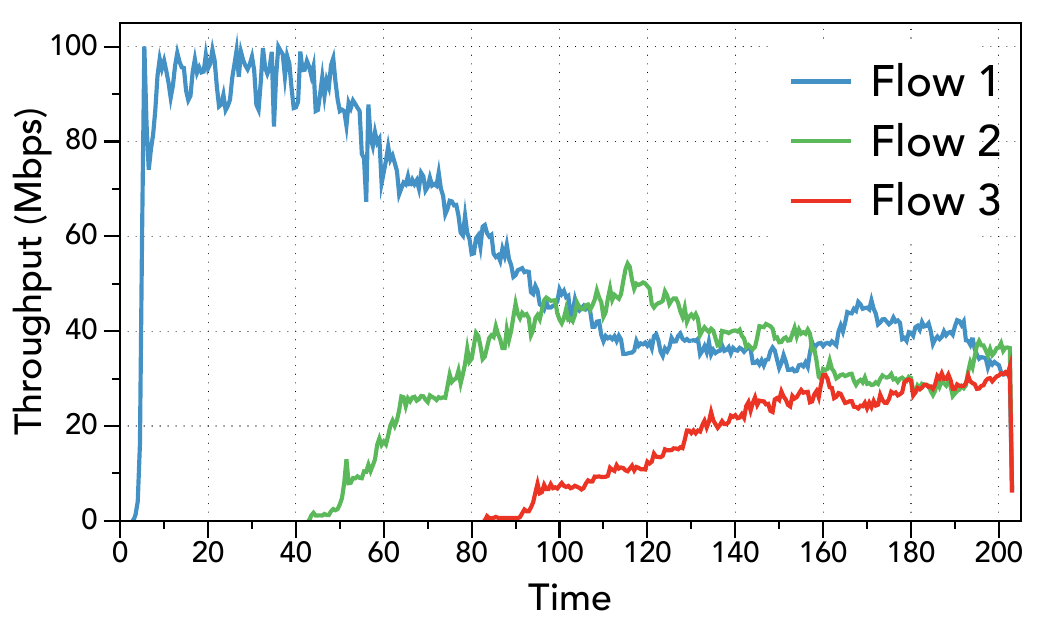}
        \caption{Vivace converges slowly.}
        \label{fig:motiv:vivace:slow}
    \end{subfigure}
    \caption{Existing learning-based algorithms fail to fast converge to fairness with stability.}
\end{figure}

%
%
\parab{Aurora is unfair.} 
We evaluate Aurora with multiple flows on a link with 80 Mbps bandwidth, 60 ms RTT, and enough buffer (4.8 MB) to absorb the traffic. As shown in Figure~\ref{fig:motiv:aurora-unfair}, Aurora is so aggressive that it would not allow other flows to get any share of bandwidth. Aurora's reward function accounts for this behavior, as it emphasizes throughput, while being ignorant of achieving stability and equal-sharing of the network bandwidth. 

\parab{Vivace shows slow responsiveness.} 
We emulate a network with a bandwidth of 100 Mbps, 1 BDP buffer size, and vary the base RTT to evaluate Vivace's convergence. We start three flows in sequence, with the launching interval of 40 seconds. In the experiment of Figure~\ref{fig:motiv:vivace:slow}, the base RTT is set to be 120 ms.
We find that Vivace can hardly achieve the fairness point before all flows terminate. The reason is that while Vivace has proof of fairness, its online learning-based rate control is based on trial-and-error, which costs several probing steps to find a better response and therefore leads to explicit inefficiency in large RTT networks.

\begin{figure}[t!]
    \centering
    \begin{subfigure}[b]{0.49\linewidth}
        \centering
        \includegraphics[width=\linewidth]{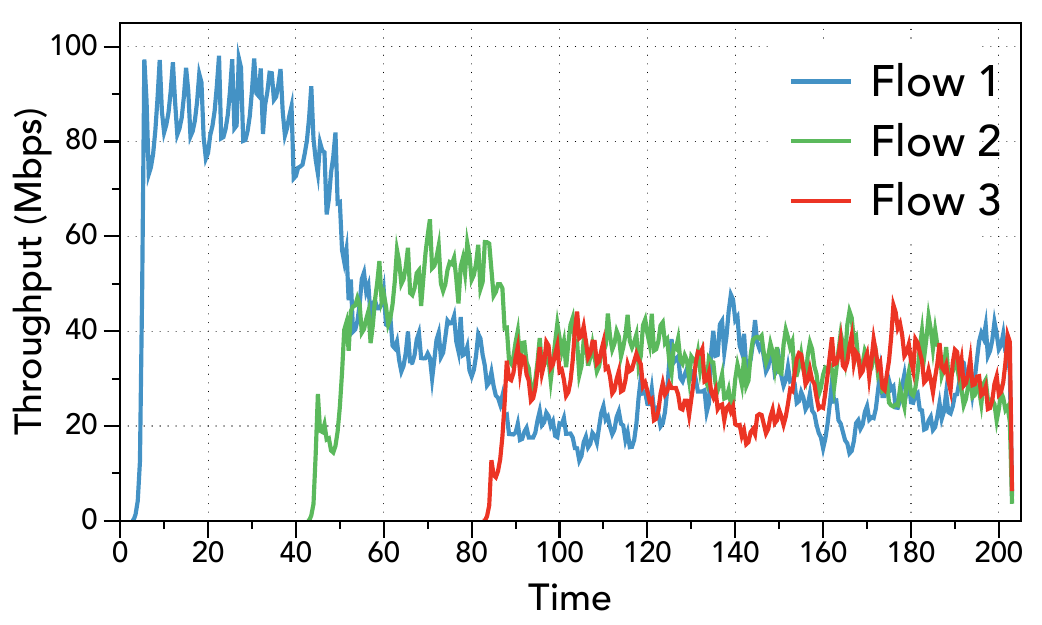}
        \caption{\footnotesize Vivace \emph{could} converge quickly.}
        \label{fig:motiv:vivace:tune}
    \end{subfigure}
    \begin{subfigure}[b]{0.49\linewidth}
        \centering
        \includegraphics[width=\linewidth]{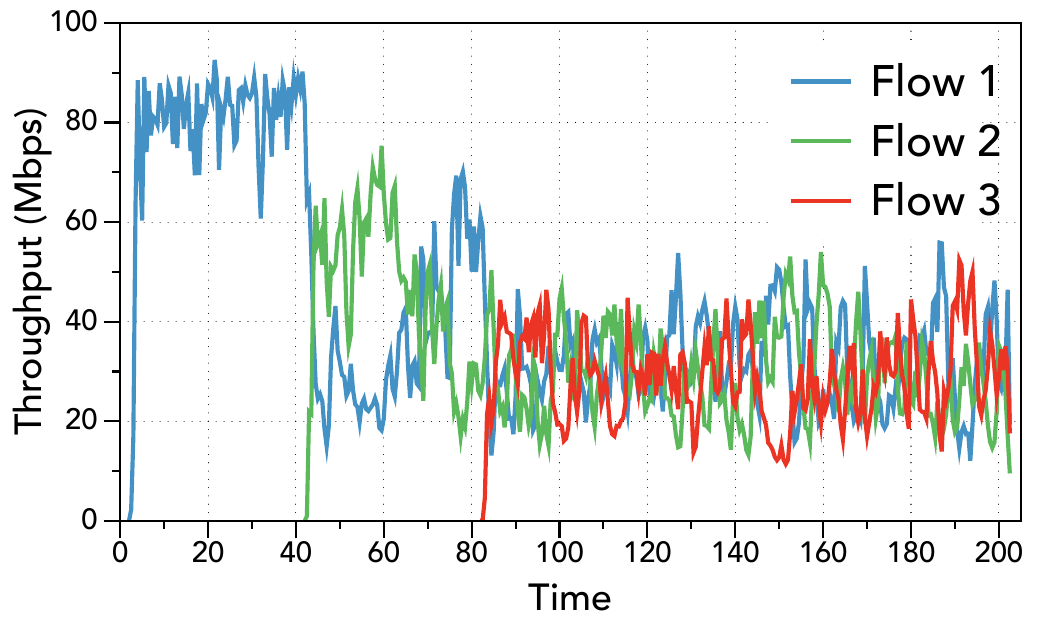}
        \caption{\footnotesize Vivace introduces instability.}
        \label{fig:motiv:vivace:instability}
    \end{subfigure}
    \caption{Enhanced Vivace performs diversely.}
\end{figure}

Both Aurora and Vivace fail to align with global properties by optimizing the local objective function, which is problematic in the real-world network environment where bandwidth and RTT may vary considerably~\cite{bbr, allegro}.  
For example, Vivace may waste the link capacity when transplanted to high-speed, \eg{} 10Gbps networks.

Furthermore, improving performance by adjusting hyperparameters on these learning-based CC schemes often leads to degradation when the network environment changes and the \textit{adjustments} become invalid again, as the mapping between the local utility function and the global goal keeps changing according to the network dynamics. In what follows, we showcase the experiment of tuning parameters in Vivace.

We attempt to improve Vivace's responsiveness by tuning its \textit{initial conversion factor}, letting Vivace put more rate increment on each probing step ($r_{new} = r + \theta_{0}\gamma)$ to converge more quickly. We properly enlarge $\theta_{0}$ and the result is presented in Figure~\ref{fig:motiv:vivace:tune}. Now Vivace becomes more responsive to other flows and can quickly lead to a fair convergence. It \textit{seems} that our adjustment aligns Vivace's local objective with global properties closer. However, as shown in Figure~\ref{fig:motiv:vivace:instability}, this will cause much instability when the \textit{enhanced} Vivace is evaluated in a network environment of 12ms RTT, where it can even hardly achieve a convergence!

To this end, we argue that existing practice of achieving these goals by optimizing local objectives is intrinsically questionable. Therefore, instead of performing the daunting work for tuning local objectives, we advocate \textit{optimizing global goals directly} for agile responsiveness to network dynamics and fast convergence to a stable and efficient state, motivating the design of \sys.


%
%
%
%

\section{Design}
\subsection{Overview}\label{sec:design:overview}
To achieve \emph{explicit} optimization on convergence properties, we design \sys to directly fulfill the insight derived in \S\ref{sec:motiv}. The major advantages of \sys come from the multi-flow training environment and the multi-agent RL algorithm that directly reward fast convergence to fairness with stability.
 
\begin{figure}[t!]
	\centering
	\includegraphics[width=\columnwidth]{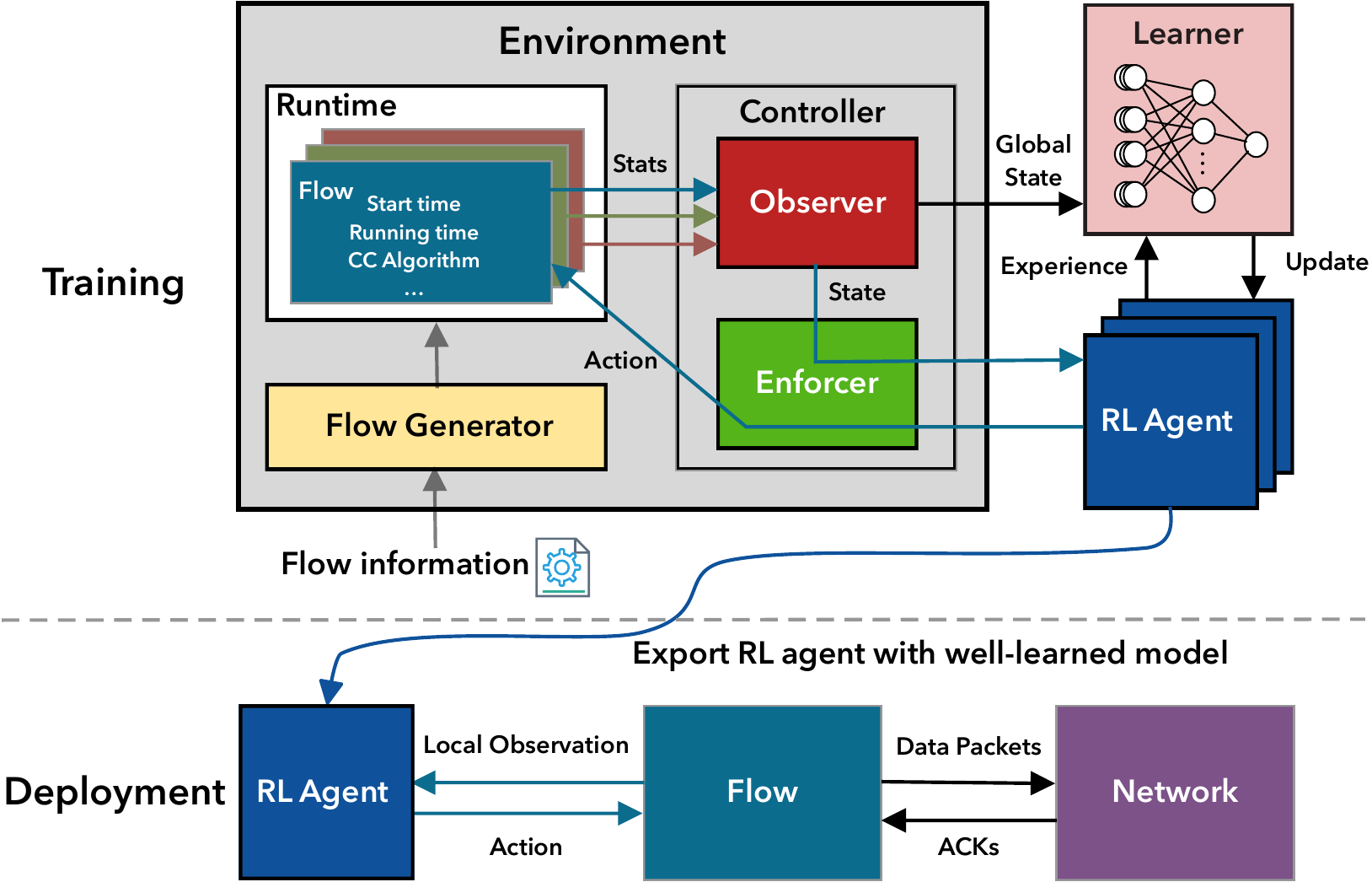}
	\vspace{1pt}
	\caption{\sys Framework.}
	\label{fig:overview}
\end{figure}

\parab{Architecture:} Figure~\ref{fig:overview} overviews \sys, which consists of three main components: a multi-flow training environment (\S\ref{sec:design:env}), RL agents which execute control policy (\S\ref{sec:design:agent}), and a Learner which updates the policy using multi-agent RL training algorithm (\S\ref{sec:design:training}). 

Basically, \sys's training environment establishes the network with multiple concurrent flows. It contains three modules: a Flow Generator, a Runtime and a Controller. With user-defined configurations, the Flow Generator starts flows runtime with pre-defined characteristics. The Controller exchanges necessary information between running flows and RL agents. Based on these modules, RL agents and the Learner in \sys perform multi-agent RL training algorithm. RL agents conduct congestion control for each flow through mapping states to actions, and the Learner collects experiences from RL agents and global information from the controller as training data to refine the policy.

\parab{Workflow:} Generally, \sys works as follows. It launches RL agents for each active flow to direct the sending behavior via exchanging information with the Controller. During training, all RL agents are initialized with a shared control policy from the centralized Learner. Each flow in the Runtime collects packet-level statistics at the granularity of Monitoring Time Period (MTP)~\cite{orca} and launches an action request to the Controller with these statistics information (local state in \S\ref{sec:design:agent}) as arguments to get the new $cwnd$. Upon receiving a request from one flow, the Observer relays the request to the Enforcer, which then forwards the request to the corresponding RL agent. At the same time, the Observer collects packet-level statistics (local state) from all other active flows to compile the global state (Table~\ref{table:global_state}) for the Learner. 
After receiving the requests from the Enforcer, the RL agents respond to it with new $cwnd$ based on the control policy, and generate interaction trajectories between flows and the Environment. The trajectories, also known as experiences in the RL area, fuel up the following RL training. 
Altogether, with the global state and the trajectories provided by RL agents, the Learner issues a global reward encoding performance and convergence metrics and trains its policy to maximize the reward objective. The updated policy is periodically pushed back to each RL agent. 

\parab{Evaluation: } In the deployment phase, as shown in Figure~\ref{fig:overview}, each flow loads one RL agent with well-learned policy from Learner to perform congestion control. Flow sender gathers local observations from each ACK and relays the information to the corresponding RL agent for further control adjustment. We note that global information is only utilized during the training process to guide the learning, and is not included as part of the model input. Therefore, the policy execution of \sys does not necessitate global information during the deployment phase.

\subsection{Environment}\label{sec:design:env}
In general, \sys environment enables the competing of multiple flows in the network and provides the interface to obtain the global states from all active participants, which is the key difference between our environment and prior solutions. Training environments in previous work only provide local information about one single flow. Therefore, they can only support direct optimization on performance metrics of throughput, latency and loss. In contrast, \sys environment sheds the light on the explicit study of global objectives, \eg convergence properties, by inferring from the global states. 

As shown in Figure~\ref{fig:overview}, within the Environment, the Runtime module consists of an emulated link with multiple concurrent flows. The link is set to simulate the network bottleneck, which is able to control the packet ingress rate and egress rate to simulate specific bandwidths and delay each packet for a certain time to simulate the base round-trip time. The emulated link can also configure user-defined queuing policies and perform randomized packet dropping to simulate non-congestion packet loss. In what follows, we demonstrate the functionalities of other modules in the multi-flow environment and illustrate the support for multi-agent training. 

\parab{Flow generator.} The Flow Generator in \sys environment starts network flows according to user-defined configuration, such as flow starting time, running time, the congestion control algorithm, and extra delay for specific flows. We introduce randomization into flow starting points and running time to emulate the dynamics of the Internet and thus enhance our environment's fidelity. In particular, we recommend modeling flow arrivals at bottleneck as Poisson arrivals~\cite{poisson-traffic} rather than deterministic arrivals, which helps RL algorithms obviate from overfitting on specific deterministic traffic patterns of network flows. The flexible configuration in our environment allows for the training and evaluation for different metrics. For instance, we can study whether flows share the bandwidth equally, by using the same CC scheme and the same base RTT for all flows (which we call homogeneous flows); also, by using various CC schemes or heterogeneous RTTs, we can study the friendliness or RTT fairness~\cite{tcp-libra} respectively.

\parab{Flow-driven control paradigm.} In a typical RL problem, the interaction between the environment and the agent is driven by the latter, which means that the RL agent proactively observes the environment, derives action from observation, and enforces the action back. However, in the network field, it's more natural and flexible to trigger the state-action control logic from the network flow side. The reason is that congestion control logic of each flow is asynchronously triggered by packet-level events or MTP expiration, which will introduce high control complexity in the agent-driven setting. Thus we design flow-side triggering where network flows proactively request control action after collecting adequate packet statistics for each MTP, making our workflow easily scalable with multiple RL agents.

\parab{Observer and Enforcer.} Controller in the environment consists of an Observer and an Enforcer. In the multi-agent RL training scenario, we need to feed the RL algorithm with observations of all active flows (we call them world observations) in the Runtime. To this end, we dedicate a centralized Observer for this purpose. When each flow sends its local packet-level statistics to the Observer to get the control action, world observation signals will be sent from the Observer to other active flows. Upon receiving signals, active flows will issue a response containing their own observations. By stacking these observations in responses, Observer then forms a global state describing the current status of the Runtime, which will be sent to the Learner for training. Enforcer performs as a proxy between the environment and RL agents: it passes the state from each flow and applies action back.


The following subsections present the design details of how \sys leverages information from the multi-flow environment to design the RL agent and the training algorithm.

\subsection{RL Agent}\label{sec:design:agent}
The RL agent decides the congestion control policy of each flow. For each monitoring time period (MTP), the agent perceives the packet statistics from the network environment and generates a new $cwnd$ for the next MTP to maximize the target performance goal. It has three key building blocks: the state block generating flow state from the world observation, the action block updating $cwnd$, and the reward block defining the performance goal. Among them, the reward block is the most challenging one, since it is difficult to measure convergence properties directly. We address this issue by carefully design a suitable representation of convergence properties from average throughput of each participant flows, which is readily computable and achievable. Besides, \sys's reward design differs from previous works~\cite{aurora, orca} by considering the packet statistics of all active flows in a centralized style, thus enabling expressing global subgoals in the network. During the training, the agent is informed with global information from these blocks to learn the optimal interaction strategy between multiple flows. We describe these blocks in detail.

\parab{State block.}
Upon receiving the packet statistics of incoming Acks in the last MTP from the world observer, the state block assembles the local and global states for the training and inference of RL agents.
The local state contains collectible statistics on end-hosts, and the global state contains global statistics across all active flows in the network.  
The state block works in two modes. In the training mode, it receives the packet statistics from all active flows in the environment, assembles both local and global states of the flow agent, and performs the multi-agent RL training algorithm. In the inference mode, only the local state is generated based on the flow information of its own and used for inference. We run the training mode for model learning and the inference mode for evaluation on the fly. 

For packet statistics received by a flow agent, we denote the throughput as $thr$, latency as $lat$, lost bytes as $loss$, the number of packet in flight as $pkt_{flight}$ and the pacing rate as $p_{rate}$. Our RL agent utilizes the following statistics in a MTP for the local state.
\begin{itemize}[leftmargin=*]
	\item The throughput ratio, the ratio of the average throughput to the maximum throughput in the flow's history $\frac{thr}{thr_{max}}$.
	\item The maximum throughput $thr_{max}$ in the flow's history.
	\item The latency ratio, the ratio of the average latency to the minimum latency in the flow's history $\frac{lat}{lat_{min}}$.
	\item The minimum latency $lat_{min}$ in the flow's history.
	\item The relative congestion control window, the current congestion control window $cwnd$ divided by the maximum throughput and the minimum latency: $\frac{cwnd}{thr_{max} * lat_{min}}$.
	\item The ratio of average loss rate to the maximum throughput $\frac{loss}{thr_{max}}$. 
	\item The ratio of the packets in flight to the current $\frac{pkt_{flight}}{cwnd}$.
	\item The ratio of the pacing rate to the maximum throughput in the flow's history $\frac{p_{rate}}{thr_{max}}$.
\end{itemize}

In order to generalize to different network conditions and accelerate training, we normalize all the features in the local state, except for the maximum throughput and minimum latency, so that the agent will have similar states that are independent to network conditions. Besides, the maximum throughput and minimum latency features will help the agent make discriminative decisions according to the network characteristics. For example, the network with high RTT links tends to respond to the agent's action more slowly, thus calling for a more conservative sending rate policy to avoid bufferbloat.

For the global state, instead of directly concatenating the local states of all active flows together, the state block performs aggregation on the local statistics vectors to reduce the input feature dimension, leading to more effective training. We calculate the minimum, maximum, and average metrics to help the agent estimate the fairness across flows. We also collect the link information, including the delay, buffer size, and bandwidth, to enable a better value estimation of the current state for the training algorithm in Section \ref{sec:design:training}. 
The full definition of global state information is shown in Table \ref{table:global_state}. These values will be used as the part of input for the critic in the training algorithm for better value estimation.

\begin{table}[ht!]
    \small
    \centering
    \fontsize{8}{10}\selectfont
    \begin{tabular}{l|l}
    \toprule
    $ovr\_thr$ & The overall throughput of all active flows. \\
    \hline
    $min\_thr$ & The minimum current throughput of all active flows.  \\
    \hline 
    $max\_thr$ & The maximum current throughput of all active flows. \\
    \hline
    $avg\_lat$ & The average latency of all active flows.  \\
    \hline 
    $min\_cwnd$ & The minimum current $cwnd$ of all active flows. \\
    \hline 
    $max\_cwnd$ & The maximum current $cwnd$ of all active flows. \\
    \hline 
    $avg\_cwnd$ & The average current $cwnd$ of all active flows. \\
    \hline 
    $loss\_ratio$ & The average loss ratio of all active flows. \\
    \hline
    $num\_flow$ & The number of active flows. \\
    \hline
    $d_0$ & The base one-way-delay of the link. \\
    \hline
    $buf$ & The buffer size. \\
    \hline
    $c$ & The bandwidth of the link. \\
    \bottomrule
    \end{tabular}
    \caption{Global state information generating by the state block.}
    \label{table:global_state}
\end{table}

To provide sufficient information for the agent to make the proper decision, the state block stacks a fixed-length (denoted by $w$) history of the per-MTP state as the final input state for the RL model, which consists of the network statistics in the last $w$ MTPs.

\parab{Action block.}
The action block returns the $cwnd$ for one flow to conduct congestion control. The action block feeds the RL model with the input state from the state block and gets the output action $a$, which is in the range $(-1,1)$. The output action $a$ is then transformed to obtain the $cwnd$ for the next MTP. We use the same mapping function between the $cwnd$ and $a$ as that in Aurora \cite{aurora}, as it controls $cwnd$ robustly and stably. With the present $cwnd_{t}$ and the output action $a_t$ in the $t$-th MTP, the action block calculates the $cwnd$ for the next MTP as follows:
\begin{equation}\label{eq:action}
	cwnd_{t+1}= \begin{cases}cwnd_{t} *\left(1+\alpha a_{t}\right) & a_{t} \geq 0 \\ cwnd_{t} /\left(1-\alpha a_{t}\right) & o.w.\end{cases}
\end{equation}   
Based on the updated $cwnd$, the pacing rate is obtained by dividing the present $cwnd$ by the smoothed RTT of packets $\frac{cwnd}{sRTT}$. In the equation \ref{eq:action}, the coefficient $\alpha$ controls the responsiveness of the agent. With a larger $\alpha$, the agent is able to exploit a larger range near the present $cwnd$ in a single MTP, which, however, may also cause an unstable sending rate and network fluctuations.  

\parab{Reward block.}
The reward block is the core building block of \sys, which defines our global goal of congestion control and works during the learning process. Specifically, we design the reward function $R$ as a linear combination of metrics reflecting our congestion control subgoals, including efficiency, stability, fairness, and responsiveness. 

Assume that there are $n$ active flows in the link. Fed with packet statistics, the reward block calculates each metric as follows. We employ the throughput metric as the ratio of the present overall throughput to the link bandwidth and the loss metric as the average ratio of the loss rate to the current throughput across flows:
\begin{equation}
R_{thr} =\frac{\sum_i{thr_i}}{c}  \qquad
R_{loss}=\frac{1}{n}\sum_i{\frac{loss_i}{thr_i}}
\end{equation}
For the latency metric, \cite{orca} proposed to ignore small queueing delay to allow flows to reach the maximum throughput in dynamic links. We employ this idea and design our latency metric as the following:
\begin{equation}
R_{lat} =\begin{cases} (\frac{1}{n}\sum_i{lat_i}-(1+\beta)d_0) p_{rate} & \frac{1}{n}\sum_i{lat_i} > (1+\beta)d_0\\ 0 & o.w.\end{cases}
\label{eq:latency}
\end{equation}
where $d_0$ denotes the base delay and $p_{rate}$ the pacing rate. We note that $R_{lat}$ will be 0 if the increased average latency is smaller than $(1+\beta)d_0$, thus small queueing size will not be penalized. Moreover, we add the pacing rate as a multiplier to penalize sending rate increment in a link that experiences high latency inflation. Thus, $R_{lat}$ can be regarded as "the total increased latency of all sending packets" in a MTP.

We measure both the fairness and stability metrics in variances of throughput, but along different axes.
Specifically, we calculate the stability metric based on the standard deviation of a flow's throughput in the fixed-length ($w$) history in the state block and the fairness metric on the standard deviation of throughputs of all active flows at the same time. We average the fairness metric along the time axis and the stability across flows to avoid the transient effect. Specifically, for fairness, instead of using the instantaneous throughput, we use flows' average throughputs in the last $w$ MTPs to calculate the standard deviation; for stability, we average the standard deviations across flows to obtain a smoother metric. 
Let $thr_{i,t}$ denotes the throughput of the $i$-th flow in $t$-th MTP. Note that these metrics are all normalized to guarantee similar rewards in various network conditions. The fairness and stability metrics are defined as follows:
\begin{equation}
\begin{split}
R_{fair} = \sqrt{\frac{\sum_i{(avg\_thr_i - \frac{1}{n}\sum_i{avg\_thr_i})^2}}{n(\sum_i{avg\_thr_i})^2}} \\
R_{stab} = \frac{1}{n}\sum_i{\sqrt{\frac{\sum_{j=0}^{w-1}(thr_{i,t-j} - avg\_thr_i)^2}{w*avg\_thr_i^2}}}
\end{split}
\end{equation}
where $avg\_thr_i$ is the average throughput of $i$-th flow in the last $w$ MTPs:
\begin{equation}
	 avg\_thr_i = \frac{1}{w}\sum{}_{j=0}^{w-1}{thr\_c_{i,t-j}}
\end{equation}
It is plain that when $R_{fair}$ and $R_{stab}$ are 0s, the flows in the network achieve the optimal fair and stability equilibriums. Putting the metrics together, we give the reward function as follows:
\begin{equation}
	R = c_0* R_{thr}-c_1*R_{lat}-c_2*R_{loss}-c_3*R_{fair}-c_4*R_{stab}
\end{equation}
We bound the reward and scale it to the range $(-0.1,0.1)$ for each MTP. With the dedicated reward function, the RL training algorithm rewards high throughput, good fairness, and stability while penalizing high latency and loss. We can adjust the coefficients $c_0,c_1,c_2,c_3,c_4$ to achieve various trade-offs between these subgoals. In general, we have tuned these coefficients to make sure that all reward terms have similar value ranges to balance their importances, and conducted an extensive search to identify a set of hyperparameter combination that performs robustly across diverse network conditions. The reward function serves as the general goal of the CC scheme under various network environments. By maintaining a consistent reward function, we ensure the RL mechanism can adaptively align the current policy with the performance objective across different scenarios.
We give the hyperparameter values in Table~\ref{table:params} in Appendix~\ref{appendix:training}. It is noted that we do not need to explicitly consider responsiveness in $R$, because RL agents are trained to maximize the accumulative reward in a horizon of future MTPs, which means that they naturally tend to reach the optimal convergence with maximum rewards as fast as possible.

\begin{figure}[t!]
    \centering
    \includegraphics[width=0.7\columnwidth]{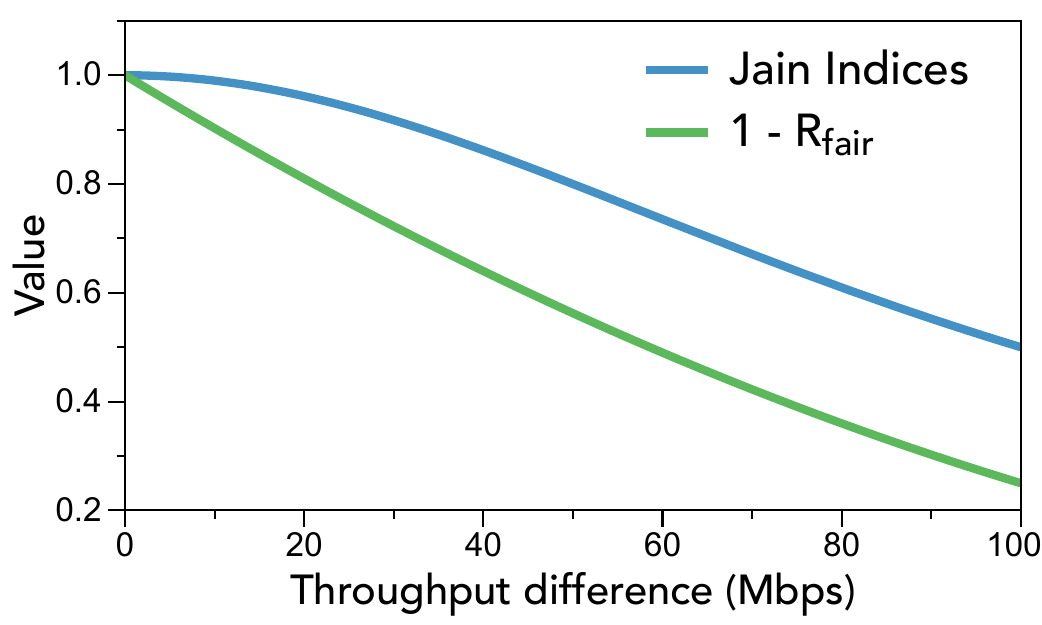}
    \caption{Jain index saturates when throughput difference of two flows approaches zero.}
    \label{fig:reward-function}
\end{figure}

\parab{Why not represent fairness metric with Jain index in training?}
While Jain index~\cite{jain} is a common metric to evaluate fairness (\S\ref{sec:exp:fair}), we find that it easily saturates when participant flows are sharing close throughput. We illustrate this phenomenon with a toy example, where there are two competing flows in one bottleneck with 100Mbps capacity and the capacity will be fully utilized by these two flows. The corresponding values across different throughput difference of these two flows are calculated in Figure~\ref{fig:reward-function}. It shows that the Jain index value is hard to differentiate two throughput with a small difference. For example, when the throughput gap of these two flows increase from 0Mbps to 20Mbps, the Jain index decreases only 0.038, comparing to 0.19 in \sys's reward\footnote{$R_{fair}$ equals zero indicates optimal fairness. Therefore we plot $1 - R_{fair}$ for easy understanding.}. Therefore, \sys's reward can preserve sensitivity for fairness when participant flows are approaching the equilibrium point, enabling \sys to keep refining its policy towards the optimal fairness.

\subsection{Multi-Agent RL Training}\label{sec:design:training}

In our multi-flow scenario, we first model the CC problem as a cooperative multi-agent reinforcement learning problem, where each agent represents a flow, and all flow agents are trained together towards a global objective. Then, we adopt a variant of the multi-agent deep deterministic policy gradient utilizing global information to solve the large variance in the multi-agent interaction environment.

\parab{Modeling} Formally, we formulate CC as a multi-agent extension of Markov Decision Process called partially observable Markov games~\cite{markovgame} with the following specific characteristics: i) it consists of multiple agents/flows interacting with the network environment and each other asynchronously; ii) the flow agents share the same policy and reward objective. At each time step $t$ for the flow $i$, the global state of the link is defined as the set of local states $(s^i_t,s^1_t,...s^{i}_t,s^{i+1}_t,...s^n_t)\in \mathcal{S}^n$ of all active flow agents as well as other useful global information. Based on the local state $s^i_t\in \mathcal{S}$, the flow agent takes action $a^i_t \in \mathcal{A}$ based on the shared policy $\pi:\mathcal{S}\rightarrow \mathcal{P}(\mathcal{A})$, where $\mathcal{P}(\mathcal{A})$ denotes the probability distribution over the action space $\mathcal{A}$. 
After executing agent's actions $a^i_t$, the state of the network changes according to a state transition function $\mathcal{T}:\mathcal{S}^n\times \mathcal{A} \rightarrow \mathcal{S}^n$ based on the network characteristics. During the state transition, the flow agent obtains a global reward $r_t$ from the environment based on the shared reward function $\mathcal{R}:\mathcal{S}^n\times  \mathcal{A} \rightarrow \mathbb{R}$. The objective of all agents is to maximize the cumulative expected return $\mathcal{J}=\mathbb{E}_{(s^1...s^n,a^1...a^n)\sim p^{\pi}}(\sum_{t=0}^{T}\gamma^t r_t)$ with finite time horizon $T$, where $p^{\pi}$ is the trajectory distribution when all flow agents follow policy $\pi$, and $\gamma$ is a discount factor.

\begin{figure}[t]
    \centering
    \includegraphics[width=0.7\columnwidth]{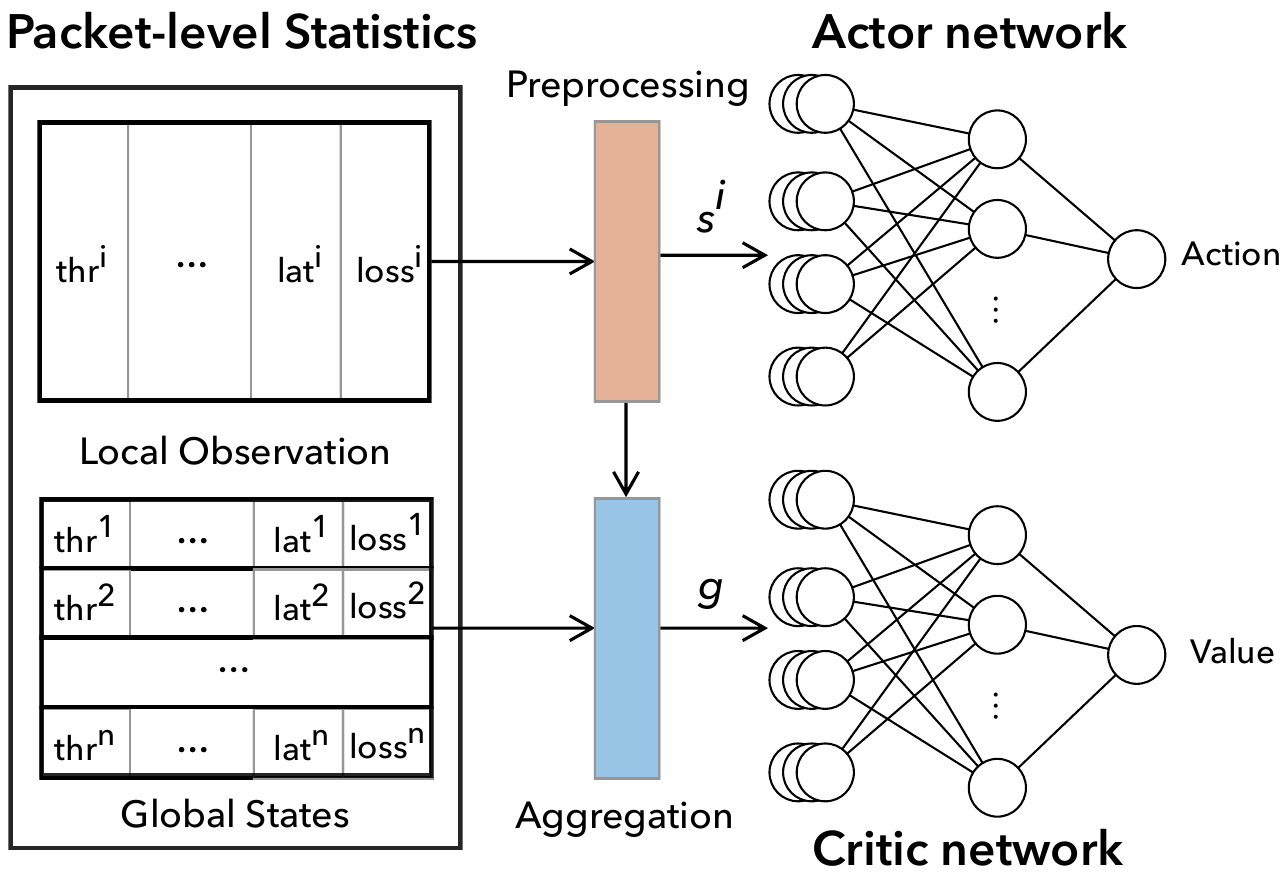}
    \caption{The model architecture in \sys's training.}
    \label{fig:network}
\end{figure}

\parab{Multi-agent actor-critic training.} 
Training in the multi-flow scenario suffers from large variance when the number of concurrent flows is large, because the reward concerning convergence depends on the interaction between all flows.  
To train the flow agents, \sys leverages a customized DRL training algorithm for multi-flow setting inspired by the multi-agent deep deterministic policy gradient (MADDPG) algorithm~\cite{marl_algorithm}.
By observing the empirical reward return from sampled trajectories, \sys calculates the gradient of the reward objective and update the policy of agents. 
The model architecture is shown in Figure \ref{fig:network}. It consists of an actor parameterizing the agent policy $\pi_{\theta}$ with $\theta$, and a critic for value estimation parameterized with $\omega$. 
In essence, the critic network functions as a reward predictor and guide, helping the actor network navigate the action space more effectively to maximize the cumulative rewards.
Given the local state $s^i$ of a flow agent $i$, the actor deterministically outputs the action value $\pi_{\theta_i}:\mathcal{S}\rightarrow \mathcal{A}$. Besides, the critic is fed with not only the agent's local state $s^i$ and action $a^i$, but also the global state $g$ aggregated from the local observations of all active flows in the state block. It approximates the action-value function $Q^{\pi_\theta}(g, s, a)=\mathbb{E}_{s^i,a^i \sim p^{\pi_{\theta}}}[\sum_{t=0}^{T}\gamma^t r_t|g,a,s]$, which is the expected future reward return from the time step when a flow agent executes action $a$ at the network state $g$. 
The technique of using extra global information to guide training in centralized learning paradigm has been employed in many previous multi-agent RL works~\cite{marl_algorithm, ActorAttentionCriticFM, qmix, MonotonicVF}. With sufficient state features, the critic provides a more precise prediction of the expected reward for the actor, which, fed with only the local link information, learns the best policy to obtain the maximum expected cumulative reward in a more stable way. During the training, we train the critic to precisely predict future reward based on the current state and action, and train the actor to output the action at that state which leads to the maximum collected reward implicated by the critic.
Both functions are approximated with deep neural networks, and the model parameters are shared across all flow agents, as they are homogeneous in the environment. 
The details of our training algorithm is as follows. To update the actor, \sys calculates the gradient of the objective function according to the deterministic policy gradient theorem \cite{ddpg} as follows:
\begin{equation}
\nabla_{\theta} \mathcal{J}\left(\theta \right)=\mathbb{E}_{s,a \sim p^{\pi_{\theta}}}\left[\nabla_{\theta} \log \boldsymbol{\pi}_{\theta}\left(a \mid s\right) Q^{\pi_\theta}\left(g,s,a\right)\right],
\end{equation}
where $Q^{\pi_\theta}(g, s, a)$ is estimated by the critic network, parameterized as $Q^{\pi_\theta}_{\omega}(g, s, a)$. By updating the actor parameters with the gradient $\nabla_{\theta} \mathcal{J}\left(\theta \right)$, the flow agents update their policy towards the direction to increase the obtained reward. The critic network is updated following the standard temporal difference method~\cite{rl_book} minimizing the objective function defined as follows:
\begin{equation}
	\small
	\mathcal{L}\left(\omega\right)=\mathbb{E}_{s, a, r, s^{\prime}}\left[\left(Q^{\pi_\theta}_{\omega}(g, s, a)-r+\left.\gamma Q^{\pi_\theta}_{\omega}(g^\prime, s^\prime, a^\prime) \right|_{a^{\prime}=\boldsymbol{\pi_\theta}\left(s^\prime\right)}
\right)^{2}\right],
\end{equation}
where $a^\prime, g^\prime, s^\prime$ denote the action, the global and local state in the following time step. It measures the relative differences between the Q values for different states and actions. By minimizing $\mathcal{L}\left(\omega\right)$ with gradient descent, the critic network is able to provide a good value estimation of $Q^{\pi_\theta}_{\omega}(g, s, a)$ for the actor to learn towards the correct direction. 

In practice, the actor and critic networks are updated based on the sampled experiences. In the training process, the state block collects and preprocesses the local observations from each agent and compiles a global state, denoted as $s^i$ and $g$. Our training algorithm receives tuples $(g,s_i,a_i,g^\prime,s_i^\prime,r)$ from the state block, calculate the gradients of $\mathcal{J}(\theta)$ and $\mathcal{L}\left(\omega\right)$, and update the actor and critic networks. The outline of \sys's training algorithm is shown in Algorithm \ref{alg:training}.

\begin{algorithm}[t]
	\small
	\LinesNumbered 
	\SetInd{0.25em}{1em}
	\caption{\sys Multi-Agent RL Training}\label{alg:training}
	\SetKwData{Left}{left}\SetKwData{This}{this}\SetKwData{Up}{up}
	\SetKwFunction{Union}{Union}\SetKwFunction{FindCompress}{FindCompress}
	\SetKwInOut{Input}{Input}\SetKwInOut{Output}{Output}
	\Input{Learning rate $\alpha$ and $\eta$, batch size $B$, episode $T$}
	\Output{Trained model parameters $\theta, \omega$}
	Initialize the actor and critic networks $\pi_\theta, Q^{\pi_\theta}_{\omega}$ \;
	\For{$t\leftarrow 0$ \KwTo $T-1$}{
	Sample $B$ experiences $(g,s_i,a_i,g^\prime,s_i^\prime,r)$ from the environment\;
	Compute the gradients for the actor:
	$\nabla_{\theta} \mathcal{J}\left(\theta \right)=\mathbb{E}_{s,a \sim p^{\pi_{\theta}}}\left[\nabla_{\theta} \log \boldsymbol{\pi}_{\theta}\left(a \mid s\right) Q^{\pi_\theta}\left(g,s,a\right)\right]$\;
	Compute the gradients for the critic:
		$\mathcal{L}\left(w\right)=\mathbb{E}_{s, a, r, s^{\prime}}\left[\left(Q^{\pi_\theta}_{\omega}(g, s, a)-r+\left.\gamma Q^{\pi_\theta}_{\omega}(g^\prime, s^\prime, a^\prime) \right|_{a^{\prime}=\boldsymbol{\pi_\theta}\left(s\right)}
\right)^{2}\right]$\;
	Update the actor and critic networks:
		$\theta\leftarrow \theta+\alpha \nabla_{\theta} \mathcal{J}\left(\theta \right), \qquad \omega\leftarrow \omega-\eta \nabla_{\omega} \mathcal{L}\left(\omega\right)$.
	}
\end{algorithm}

\parab{Online learning:} Due to the centralized training of \sys, it is non-trivial to retrain our model during the online phase, as the global information may not be available to end hosts. Furthermore, due to the distributed nature of CC, distributed online learning may lead to diversified CC schemes for each end-host, which may cause unpredictable fairness issues between competing flows. Therefore in this paper, we pre-train our CC model offline and distribute the trained model across end hosts. A possibly reasonable continuous learning method of \sys is that the network administrator periodically conducts the retraining with collected network experience and distributes the updated models to end hosts.

\section{Implementation}\label{sec:impl}

\parab{Training environment.} We build the training environment of \sys with Mahimahi~\cite{mahimahi} and Pantheon-tunnel~\cite{tunnel} in Python for compatibility, implementing the functionalities described in \S\ref{sec:design:env}. Mahimahi is used to establish the virtual link for mimicking the network bottleneck, in which every network flow runs in one virtual tunnel built on the top of \cite{tunnel} to guarantee isolation. The communication between network flows and the Controller in the environment is built with UNIX socket for low latency. In particular, we enable asynchronous action fetching and packet statistics collection to mitigate model inference overhead. 

\parab{Training scheme and hyperparameters.} For the multi-agent RL training scheme, we use the same learning rate $\alpha=0.001$ for both actor and critic networks. Both networks use 3-layer multi-layer perceptrons, with 256, 128 and 64 neurons in each layer. The model is updated for a fixed number of model update step (20 steps) every time the environment runs for the model update interval (5 seconds). We execute the training on various network conditions so that the learned agent can be generalized to unseen network environments. 
The environment characteristics are shown in Table \ref{table:condition}. More training hyperparameters are given in Appendix~\ref{appendix:training}. We assign multiple running flows in the link with different RTTs to include heterogeneity. The number of concurrent flows of each training episode is randomly sampled from 2 to 5.

\parab{\sys prototype.} Based on the multi-flow training environment and multi-agent RL algorithm, we implement a fully functional \sys prototype. We implement the sender to leverage \sys's well trained model to conduct congestion control, in which fetching information of kernel TCP flows and enforcing actions are built with custom socket options. We have implemented a customized congestion control building block for Linux kernel to bypass regular congestion avoidance from underlying TCP schemes, so that \sys can fully control the sending process.

\begin{table}[t!]
	\centering
	\begin{tabular}{c|c|c}
	\toprule 	
	Bandwidth & Base RTT & Buffer size factor \\
	\hline
	[40Mbps-160Mbps] & [10ms-140ms] & [0.1-16]\\
	\bottomrule
	\end{tabular}
	\caption{Training environment characteristics.}
	\label{table:condition}
\end{table}

\parab{\sys{} inference service.} To make \sys{} scalable with a large number of concurrent flows, we implement an \sys{} inference service which is able to serve multiple \sys{} senders. The service is implemented in C++ with TensorFlow C++ APIs for high efficiency. The communication between the inference service and \sys{} senders is built with UNIX socket or UDP socket. To reduce the system overhead, the inference service works in a batch mode, \ie it collects inference requests from multiple senders within a fixed time interval (5ms) as a batch and serves them simultaneously.


\section{Evaluation}\label{sec:exp}

In this section, with the selected benchmarks in previous work~\cite{vivace, orca}, we evaluate the performance of \sys and demonstrate its advantages over typical TCP schemes (Cubic~\cite{cubic}, Vegas~\cite{vegas} and BBR~\cite{bbr}), recent learning-based algorithms (Remy~\cite{remy}, Aurora~\cite{aurora}, Vivace~\cite{vivace}, Orca~\cite{orca}) and delay-based scheme Copa~\cite{copa} through extensive experiments in real-world Internet and emulated environments\footnote{The used models and parameters in compared schemes are sourced from their original papers.}. Our evaluation centers around several key questions:
\begin{itemize}[leftmargin=*]
	\item \textbf{How does \sys improve fairness?} We investigate convergence properties of \sys and other schemes by launching multiple flows in emulated network. Our experiments show that \sys achieves near-optimal bandwidth sharing among multiple homogeneous flows, achieves 8.4$\times$ speed up on convergence time, and delivers more stable throughputs at up to 2.8$\times$ compared with Vivace (\S\ref{sec:exp:fair:simple}). We demonstrate that \sys achieves comparable RTT fairness (\S\ref{sec:exp:fair:rtt-fair}) with existing schemes. We then show that \sys is fair with consistent high Jain Index across various network conditions (\S\ref{sec:exp:fair:diverse}) and exhibits near max-min fairness in multi-bottleneck scenarios  (\S\ref{sec:exp:fair:multi-bottleneck}). In addition, we seek to understand \sys's internal working principle to achieve fairness in \S\ref{sec:exp:hood}. We also conduct sensitivity experiment on fairness coefficient, and find that \sys can preserve consistently high Jain index with a range of coefficient from 0.05 to 0.35 in Appendix~\ref{appendix:training}.
	\item \textbf{How does \sys improve convergence speed and stability?} Our experiments show that \sys achieves 8.4$\times$ speed up on convergence time, and delivers more stable throughputs at up to 2.8$\times$ compared with Vivace (\S\ref{sec:exp:conv}) In addition, we illustrate the quick responsiveness of \sys in cellular network environments (\S\ref{sec:exp:conv}).
	\item \textbf{How does \sys maintain high performance in emulation and real-world experiments?} First we show that \sys exhibit comparable TCP friendliness and is not vulnerable to be inefficient when competing with Cubic flows (\S\ref{sec:exp:perf:friend}). In addition, real-world experiments (\S\ref{sec:exp:perf:real-world}) show that, though focusing on convergence properties, \sys always defines the frontier that shows high throughput with moderate latency. For instance, \sys achieves 1.4$\times$ lower latency inflation than BBR and 3.1$\times$ high throughput than Orca. 
	Besides, we present the low CPU overhead of \sys in \S\ref{sec:exp:overhead}. We provide additional experimental results, including \sys{}'s performance in varying buffers, satellite networks and high-speed networks, in Appendix~\ref{sec:exp:emu}
\end{itemize}

 

\subsection{Fairness}\label{sec:exp:fair}

\subsubsection{Fairness of Multiple Homogeneous Flows}\label{sec:exp:fair:simple}
To understand how \sys responds when flows arrive and leave, we emulate a link with 100 Mbps bandwidth with 30ms RTT and 1 BDP buffer. We start 3 flows at the interval of 40s; each flow runs for 120s so that flows co-exist with adequate time length. We omit the results of Aurora in this part since we have shown its unfairness in \S\ref{sec:motiv}.

\begin{figure}[t!]
    \centering
    \begin{subfigure}[b]{0.49\columnwidth}
        \centering
        \includegraphics[width=\linewidth]{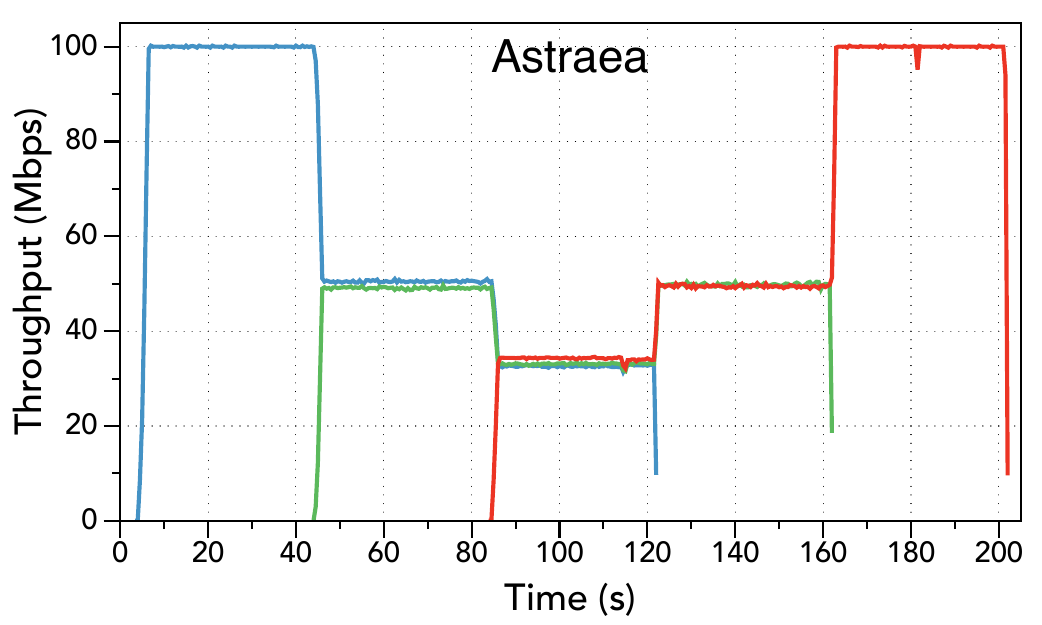}
    \end{subfigure}
    \begin{subfigure}[b]{0.49\columnwidth}
        \centering
        \includegraphics[width=\linewidth]{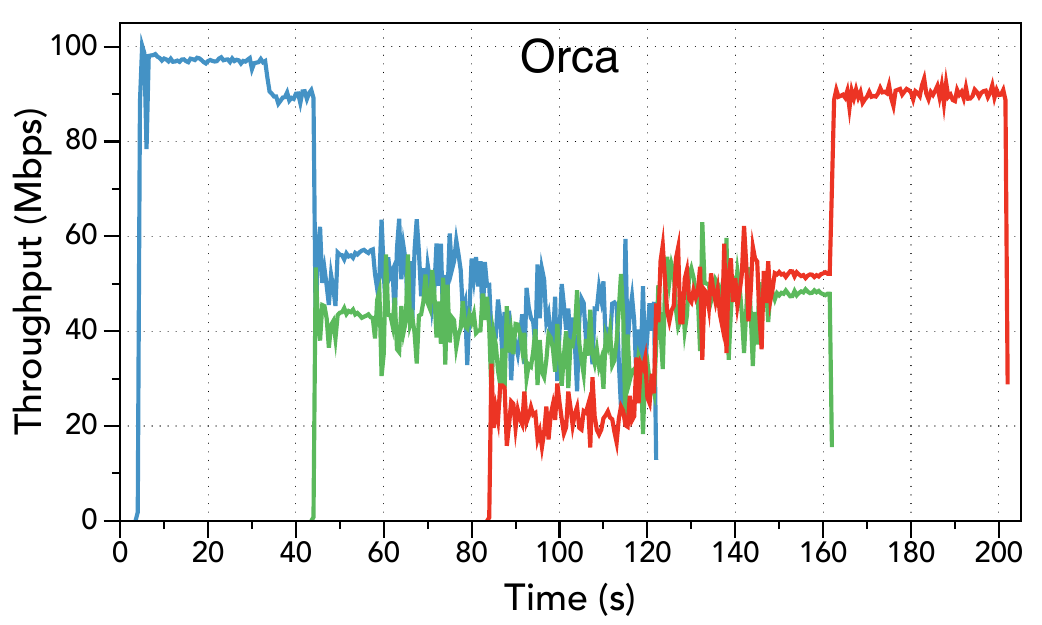}
    \end{subfigure}
    \begin{subfigure}[b]{0.49\columnwidth}
        \centering
        \includegraphics[width=\linewidth]{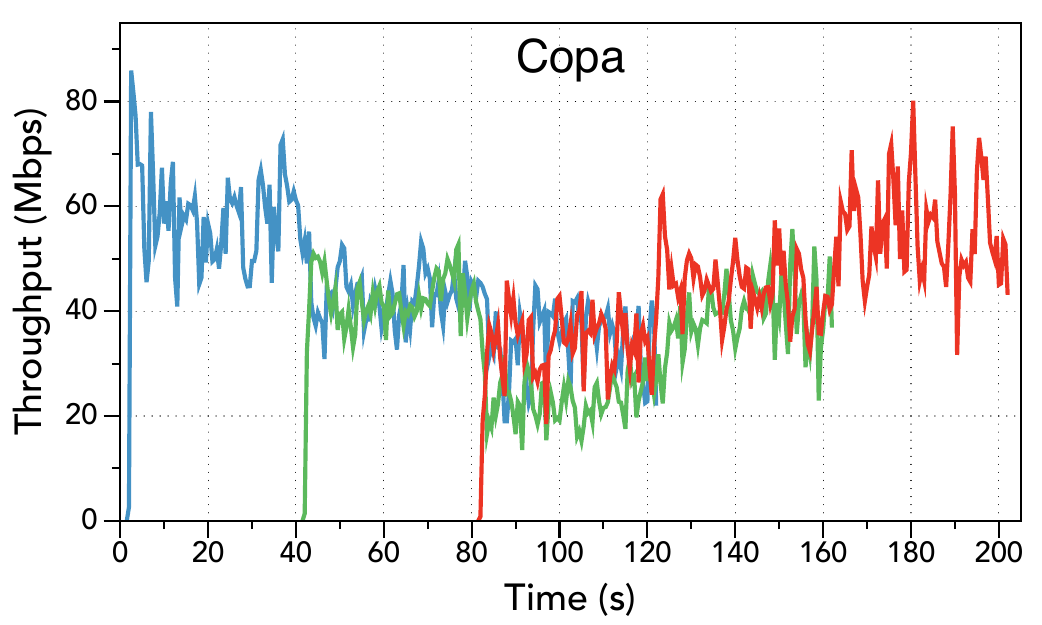}
    \end{subfigure} 
    \begin{subfigure}[b]{0.49\columnwidth}
        \centering
        \includegraphics[width=\textwidth]{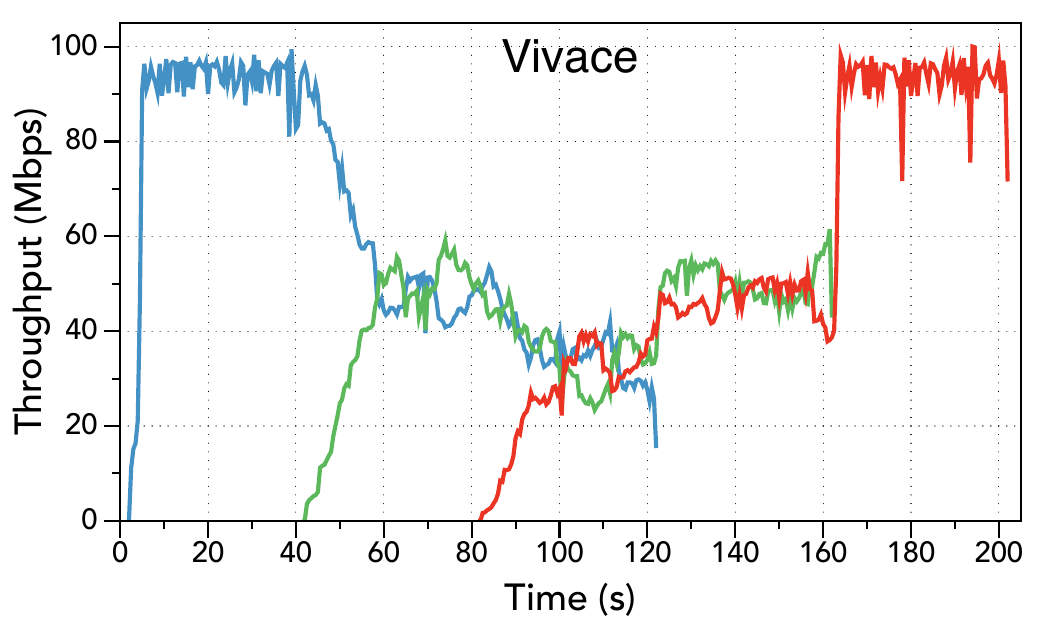}
    \end{subfigure}
    \begin{subfigure}[b]{0.49\columnwidth}
        \centering
        \includegraphics[width=\linewidth]{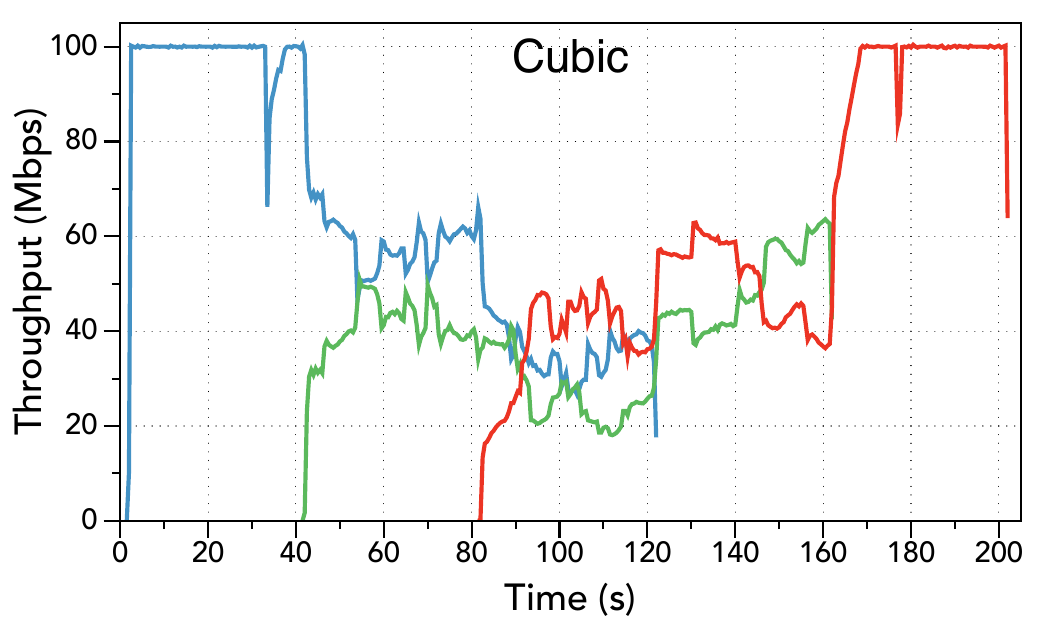}
    \end{subfigure} 
    \begin{subfigure}[b]{0.49\columnwidth}
        \centering
        \includegraphics[width=\textwidth]{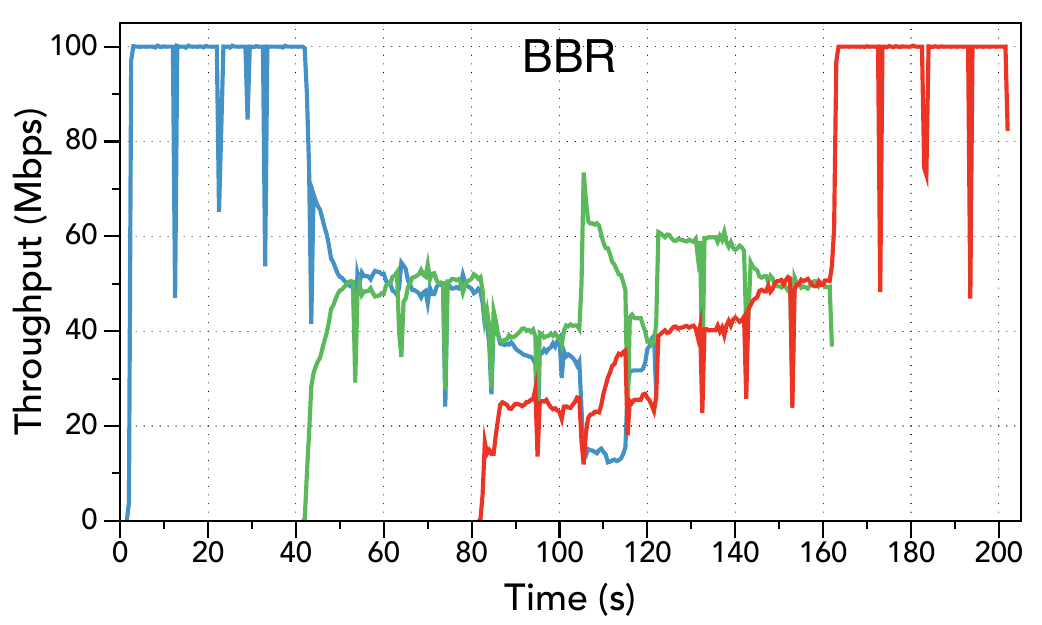}
    \end{subfigure}
    \begin{subfigure}[b]{0.49\linewidth}
        \centering
		\includegraphics[width=\columnwidth]{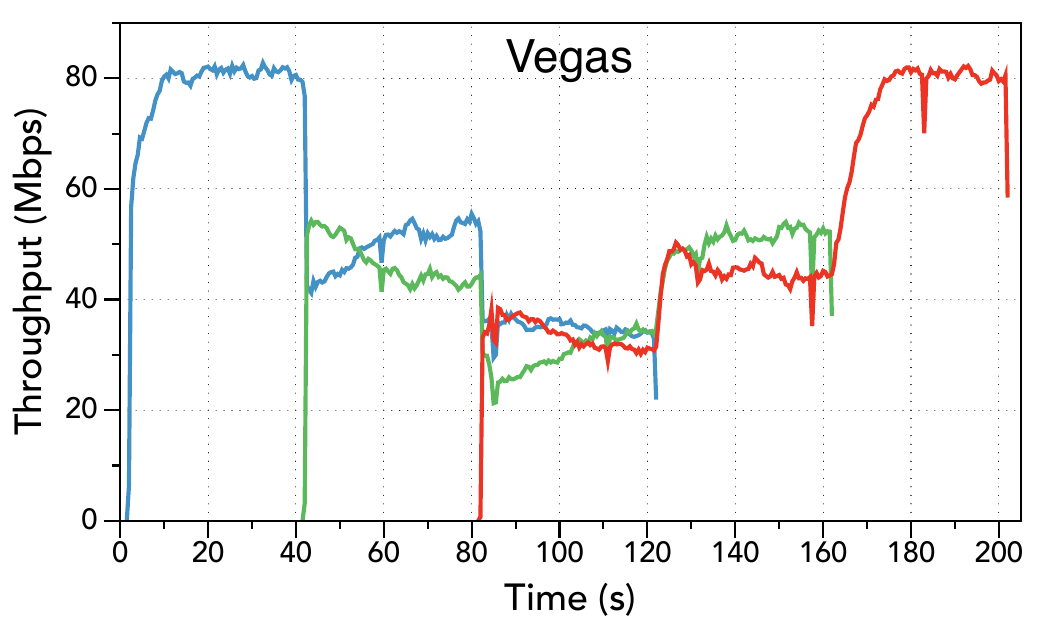}
    \end{subfigure}
    \begin{subfigure}[b]{0.49\linewidth}
        \centering
        \includegraphics[width=\linewidth]{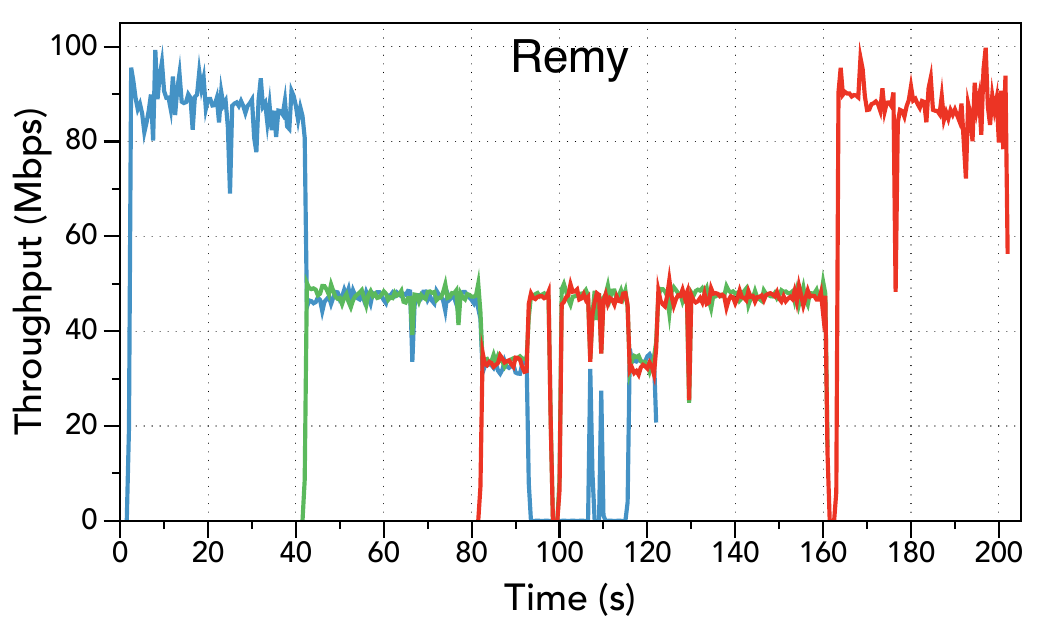}
    \end{subfigure}
    \caption{Temporal behavior of convergence of various congestion control schemes.}
    \label{fig:exp:fair}
\end{figure}

We first report the convergence process of each scheme in Figure~\ref{fig:exp:fair}. In general, all TCPs can quickly respond to flow arrivals and departures. Cubic and BBR achieve high link utilization while introducing significant rate oscillations. 
Also, the delay-based scheme, Copa, can quickly respond to flow events, yet demonstrates significant instability, which may due to the erroneous switches to the competitive mode for a few RTTs~\cite{copa}. It is interesting to observe that Vivace exhibits relatively slow responsiveness compared with all other schemes, as it needs to perform probing steps before deciding the direction to change rate. Orca mitigates instability compared to its default underlying scheme Cubic. However, its convergence is still far from optimal. The reason is that even if Orca learns to act conservatively to control the rate oscillation issue of its underlying Cubic, it can hardly achieve satisfactory stable fairness through implicit optimization, because the fairness metric is not included in its local reward function. Compared with all these schemes, \sys exhibits near-optimal fast convergence with efficiency and high stability. The reason is two folds: 1) \sys explicitly includes convergence metrics in optimization goals and performs efficient training; and 2) it entirely controls the sending behavior, avoiding the variance from underlying TCP schemes. 

\begin{figure*}[t]
    \centering
    \begin{minipage}[t]{0.32\textwidth}
        \centering
        \includegraphics[width=\textwidth]{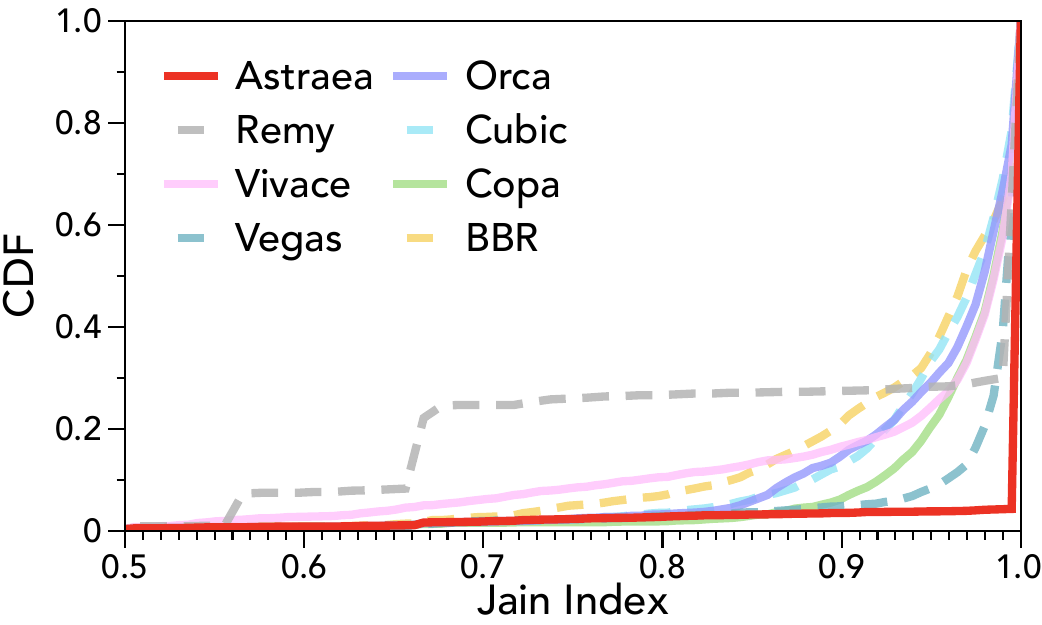}
        \caption{CDF of Jain indices calculated at various timeslots for multiple flow experiments.}
        \label{fig:exp:fairness:jain}
    \end{minipage}
    \hfill 
    \begin{minipage}[t]{0.32\textwidth}
        \centering
        \includegraphics[width=\textwidth]{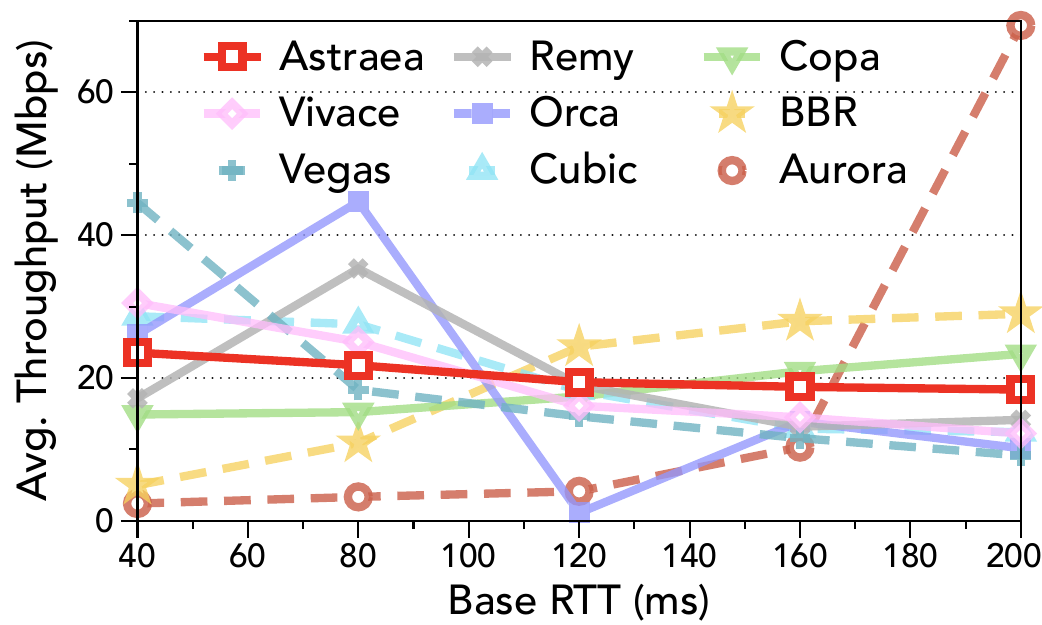}
        \caption{RTT-fairness of various congestion control schemes. 20Mbps throughput means optimal sharing.}
        \label{fig:exp:rtt-fairness}
    \end{minipage}
    \hfill
    \begin{minipage}[t]{0.32\textwidth}
        \centering
        \includegraphics[width=\textwidth]{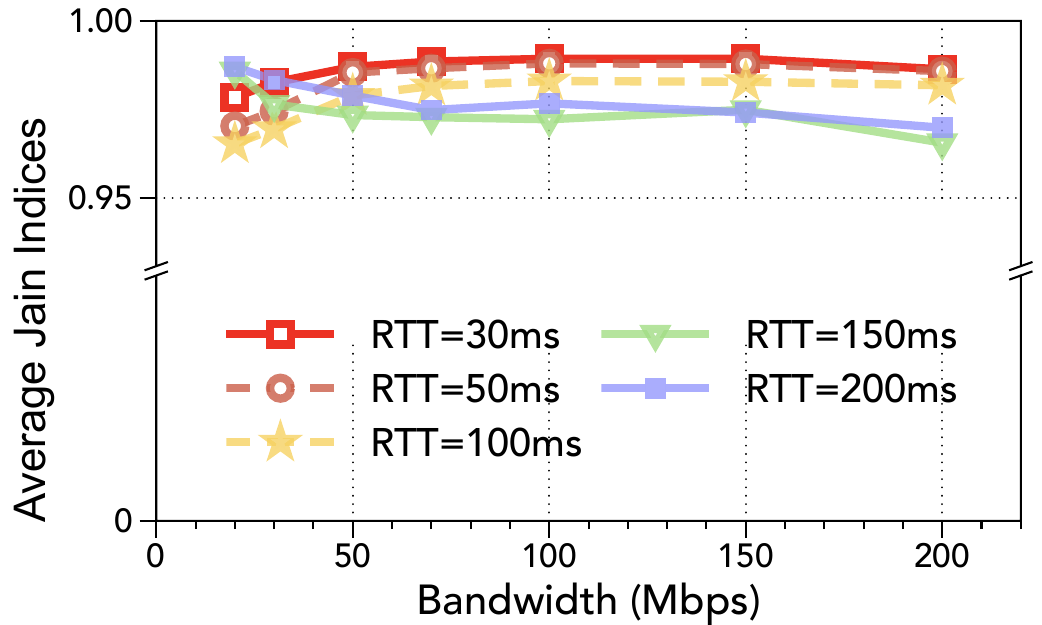}
        \caption{Jain indices in diverse network scenarios.}
        \label{fig:exp:vari-jain-index}
    \end{minipage}
\end{figure*}


We then repeat each test 10 times and report the statistical values to describe the fairness property of \sys. The overall CDF of Jain Index is depicted in Figure~\ref{fig:exp:fairness:jain}. The Jain index is calculated in all the timeslots that have at least two active flows. One key observation is that \sys achieves almost the full Jain Index all the time, \ie active \sys flows can always equally share bandwidth with incoming flows quickly. The distribution of \sys's Jain Index solidifies the intuition of including convergence properties explicitly in the optimization objective.

\subsubsection{RTT Fairness}\label{sec:exp:fair:rtt-fair}
In addition to studying the competing behaviors of multiple flows that possess the same RTT, we further inspect how \sys performs when there are multiple concurrent flows with different RTTs. Ideally, flows sharing the same bottleneck should get identical throughputs. To experimentally evaluate the RTT fairness, we set up 5 long-running flows in an emulated link with 100Mbps bandwidth. The base RTT of each flow evenly spaces between 40ms and 200ms. The link has 1 BDP buffer that calculates with 200ms RTT. We repeat each test 10 times. The average obtained throughput is reported in Figure~\ref{fig:exp:rtt-fairness}.

We observe that \sys exhibits milder throughput distance to the optimal, showing comparable RTT fairness with Copa and Vivace, and outperforming Aurora, Orca and other TCP schemes. The reason is two folds: 1) We have introduced the RTT heterogeneity in our training setting to explicitly improve \sys's RTT fairness; and 2) The CWND adjustment in \sys is RTT independent. As presented in \S\ref{sec:design:agent}, the fairness reward signal in \sys framework does not differentiate RTTs, thus poses the same advantages over them. With this reward logic, multiple heterogeneous flows turn to achieve fair sharing to maximize the incentive of \sys's objective. However, \sys still exhibits some unfairness when the RTTs are small. This is because \sys essentially operates as a time-interval based algorithm. In small RTT scenarios, \sys gets faster feedback from the network and therefore responds more swiftly, which leads to its throughput advantage.



\subsubsection{Fairness in More Diverse Network Scenarios}\label{sec:exp:fair:diverse}

To understand convergence properties of \sys in more diverse network scenarios, we establish a bottleneck with changing bandwidth and base RTTs, and run multiple \sys flows inside the bottleneck to observe the degrees of bandwidth fair sharing. The bandwidth and base RTT range from 20Mbps to 200Mbps and 30ms to 200ms respectively (wider than the training range), which we believe can fill the basic range of the Internet. The number of flows of each trial is randomly sampled from 2 to 8. We set the flow starting interval as 20 seconds and tune the running time of each flow accordingly to guarantee the adequate competing similar to \S\ref{sec:exp:fair:simple}. 

We report the convergence results by plotting the average Jain indices in Figure~\ref{fig:exp:vari-jain-index}. Each point in this figure presents the average Jain Index of 10 trials under one specific network configuration. Overall, we find that \sys achieves high jain indices across all evaluated network scenarios (higher than 0.95), which demonstrates that \sys can preserve good fairness across a wide range of network conditions. Furthermore, \sys provides decent fairness in large RTT scenarios (\eg 200ms) beyond its training range, which indicates that it can generalize to unseen network conditions.

Besides, we make two additional observations of \sys's fairness under this wide range of network scenarios: 1) The Jain Index of \sys degrades under large RTT scenarios. The reason comes from the intrinsic slow feedback from the network, which makes it hard for \sys to perceive the co-existence of other flows in a timely manner. Therefore, \sys may perform rate occupation and relinquishing sluggishly and results in slow convergence speed to fairness point. Thus, \sys delivers moderate fairness under network conditions of 150ms and 200ms RTTs. 2) The \sys fairness performance may degrade in small BDP networks (low bandwidths will small RTTs). For example, there is a performance gap between (50Mbps, 30ms) scenario and (20Mbps, 30ms) conditions. The degradation may be due to the action formulation of \sys, which will lead to rounding error when the CWND is small, as only integer output is supported.

\begin{figure}[t]
    \centering
	\includegraphics[width=0.8\columnwidth]{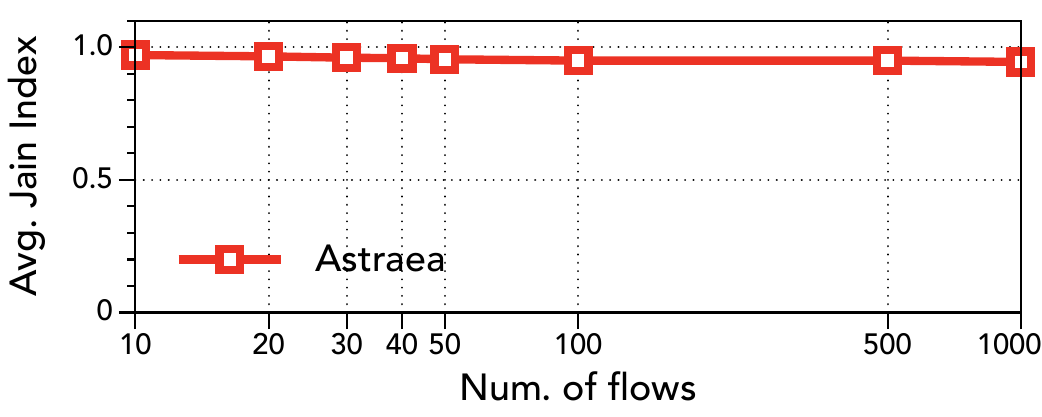}
	\caption{Fairness of varying number of competing flows.}
	\label{fig:exp:fair:large-flow}
\end{figure}

Given that there may be many concurrent flows over the backbone network in Internet~\cite{sizing}, we seek to evaluate \sys's fairness in the scenario of a large number of competing flows.
We establish an emulated bottleneck with 600Mbps bandwidth and 20ms RTT and increase the number of competing flows in the bottleneck from 10 to 50 (the maximum number of flows supported by our emulated environment). We also extend this experiment with a larger number of competing flows (up to 1000) in a link using Linux TC qdisc~\cite{tc}. Figure~\ref{fig:exp:fair:large-flow} reports the average Jain Index of 10 trials under different numbers of competing flows. It shows that \sys preserves a high degree of fairness by exhibiting high Jain indices across all evaluated scenarios, though it is trained with a limited number of flows. We attribute the good generalizability of \sys's fairness property to our normalization techniques, and \S\ref{sec:exp:hood} gives a more illustrative explanation.


\subsubsection{Fairness in Multi-bottlenecked Scenarios}\label{sec:exp:fair:multi-bottleneck}
\begin{figure}[t!]
    \centering
    \begin{subfigure}[b]{0.43\linewidth}
        \centering
        \includegraphics[width=\linewidth]{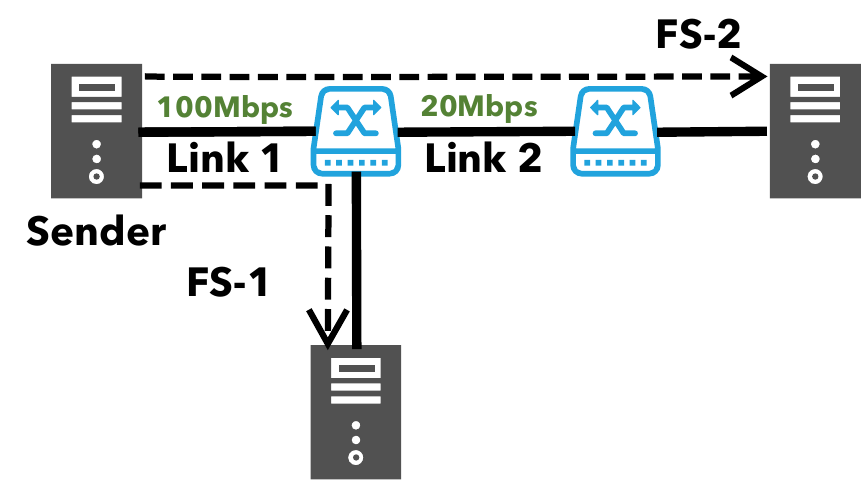}
        \caption{Topology.}
        \label{fig:exp:multi-bottleneck:topo}
    \end{subfigure}
    \begin{subfigure}[b]{0.55\linewidth}
        \centering
        \includegraphics[width=\linewidth]{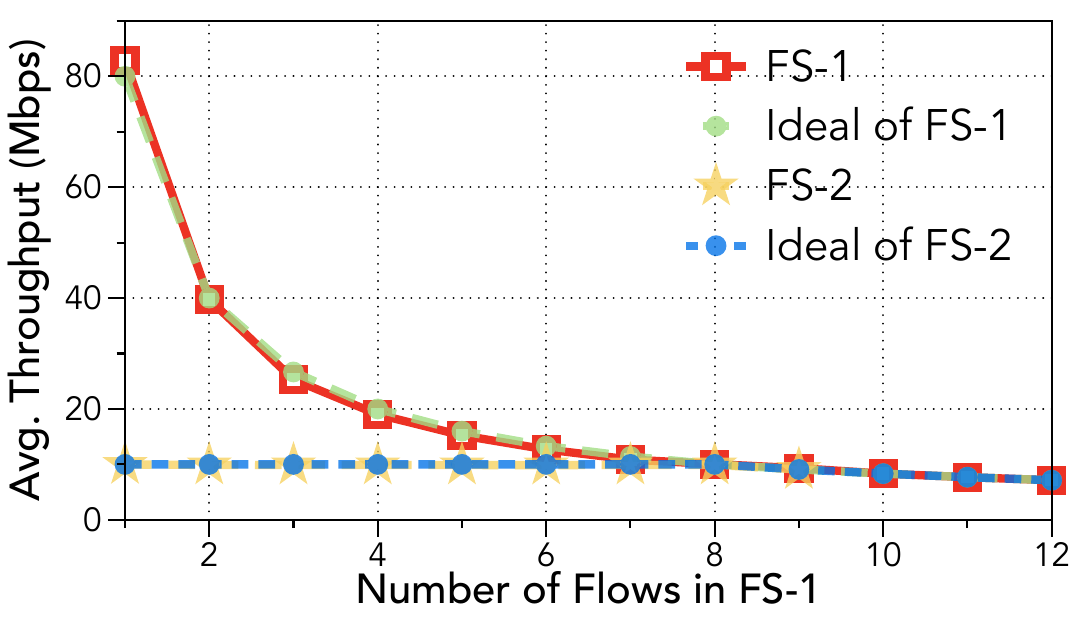}
        \caption{Throughput of FS-1 and FS-2.}
        \label{fig:exp:multi-bottleneck:tput}
    \end{subfigure}
    \caption{Fairness in multi-bottleneck topology.}
\end{figure}
To further show \sys's fairness properties in the scenario of multiple bottlenecks, we set up a topology in Figure~\ref{fig:exp:multi-bottleneck:topo} following the same multi-bottleneck setting in~\cite{expressPass}, in which Flow set 1 (FS-1) and Flow set 2 (FS-2) (representing two sets of flows) share the common Link 1. We set Link 1 and Link 2 to use 100Mbps and 20Mbps bandwidth, respectively, both of which use the 30ms base RTT with adequate buffers. We establish two flows in FS-2 and vary the flow numbers of FS-1. We start FS-1 and FS-2 simultaneously and present their average throughputs of 10 trials in Figure~\ref{fig:exp:multi-bottleneck:tput}. 

We observe that both the average throughputs of FS-1 and FS-2 closely follow the ideal cases. When the flow number of FS-1 is below 8, its bottleneck link is Link 1 and FS-2's bottleneck is Link 2. As a result, two flows in FS-2 fairly share Link 2's 10Mbps and all flows in FS-1 equally share the remaining 80Mbps in Link 1. When the flow number of FS-1 further increases, Link 1 becomes the common bottleneck for FS-1 and FS-2. In this case, all flows fairly share the 100Mbps bandwidth. The reason why \sys{} can deliver good fairness in the multi-bottleneck scenario is that it has been trained to ensure fairness under varying bandwidths and RTTs, which is in fact analogous to the alterations of sending rates in irrelevant flows in multi-bottleneck scenarios. Therefore, the sending rate changes of irrelevant flows will not affect the fairness among flows sharing the bottleneck.

\subsection{Convergence Speed and Stability}\label{sec:exp:conv}
\begin{figure*}[ht!]
    \centering
    \begin{minipage}[t]{0.32\textwidth}
        \centering
        \includegraphics[width=0.8\columnwidth]{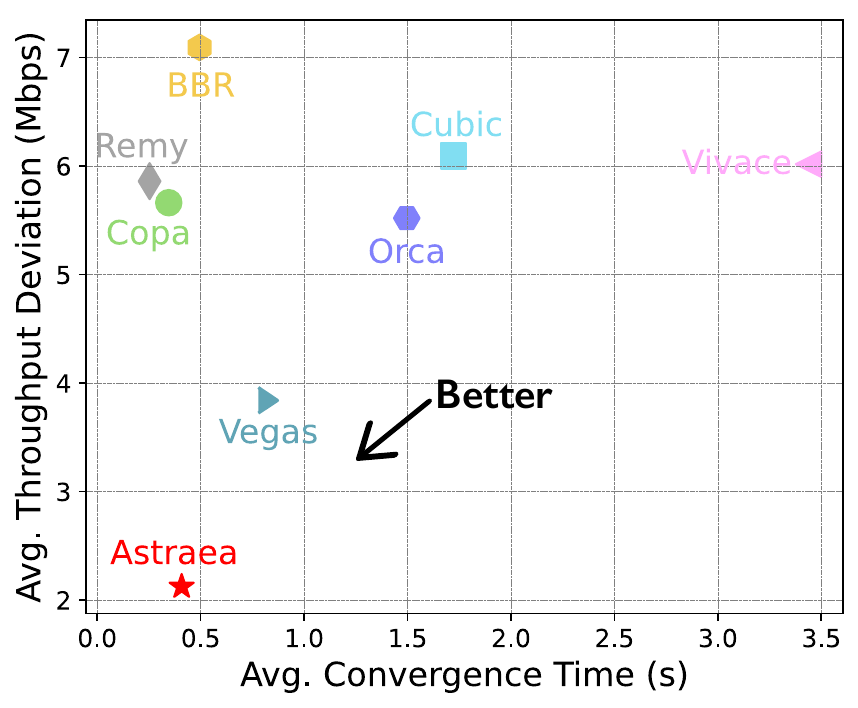}
        \caption{\small Convergence Time vs Stability.}
        \label{fig:exp:fairness:conv}
    \end{minipage}
    \hfill 
    \begin{minipage}[t]{0.32\textwidth}
        \centering
        \includegraphics[width=\columnwidth]{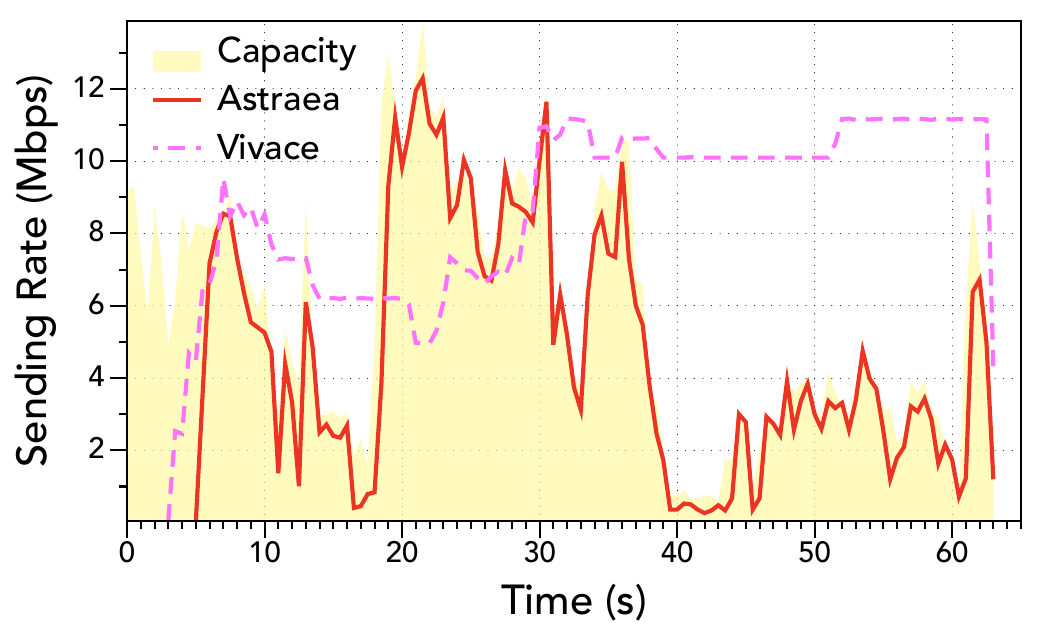}
        \caption{\sys can adapt to rapidly changing cellular network conditions.}
        \label{fig:exp:cellular-tput}
    \end{minipage}
    \hfill
    \begin{minipage}[t]{0.32\textwidth}
        \centering
        \includegraphics[width=\columnwidth]{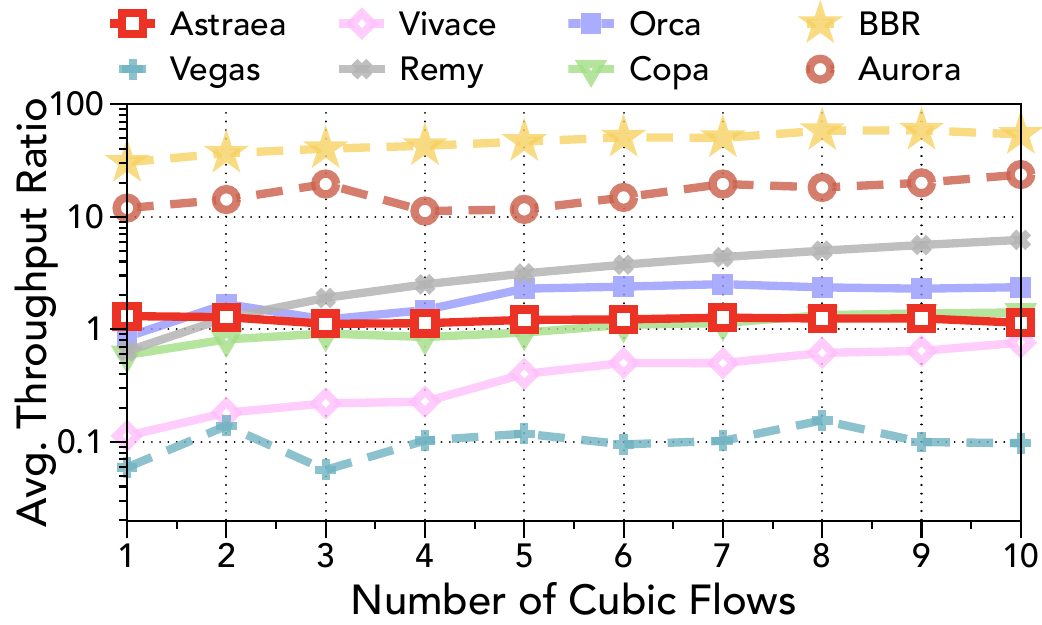}
        \caption{Throughput ratio to CUBIC of various CC schemes. A ratio of 1 indicates optimal friendliness.}
        \label{fig:exp:friendliness}
    \end{minipage}
\end{figure*}
To demonstrate the convergence process numerically in multi-flow competing scenario, we report the convergence time and stability of experiments in \S\ref{sec:exp:fair:simple}. We calculate the convergence time as the time from flow events (flow arrival or departure) to the time when it reaches a sending rate within $\pm$10\% of its ideal fair share. 
The convergence stability is calculated as the standard deviation of the throughput of the new arrival flow after its convergence.
As shown in Figure~\ref{fig:exp:fairness:conv}, \sys achieves the average convergence time of 0.408s, which is comparable with Copa, 3.7$\times$ faster than Orca (1.497s), and 8.4$\times$ faster than Vivace (3.438s), respectively. In addition, \sys achieves the best stability (2.124 Mbps) among all evaluated CC schemes, which is 2.6$\times$ better than Orca (5.519 Mbps), and 2.8$\times$ better than Vivace (6.016 Mbps).

To further understand the responsiveness of \sys, we evaluate it in a changing network environment. Cellular wireless networks are tricky and challenging for congestion control as their link speed varies drastically in the matter of milliseconds, which requires CC scheme to be agile to bandwidth variances and resilient to random noise. We use Mahimahi to replay the LTE trace provided by \cite{sprout} and compare \sys with other schemes in terms of average throughput and normalized latency. The emulated link has 40 ms RTT and a very deep buffer to absorb the traffic. We repeat the test on each CC scheme 10 times.

We first compare the sending rate dynamics of \sys with Vivace in Figure~\ref{fig:exp:cellular-tput}.
As shown in the results, \sys can adjust sending rate swiftly to align with the changes of link capacity, which demonstrates the responsiveness. On the other hand, Vivace cannot capture the changes of link capacity and respond with improper rate adjustment, which therefore causes extreme latency inflation and severe packet losses. The reason is that Vivace's probe-and-decide control logic slows down its reaction to the rapidly changing capacity. It is noted that \sys is not specially trained for this kind of network environment. More statistical results are presented in Appendix~\ref{sec:exp:resp:changing}.


\subsection{High Performance}\label{sec:exp:perf}
\subsubsection{TCP Friendliness}\label{sec:exp:perf:friend}

We study how \sys behaves when competing with Cubic flows to show its friendliness. We create an emulated link with 100Mbps bandwidth, 30ms RTT and 1 BDP buffer size, and start one evaluated flow with the increasing number of Cubic flows. 
We repeat each trial 10 times. Figure~\ref{fig:exp:friendliness} depicts the average ratio of throughput achieved by the evaluated scheme and the average throughput of other Cubic flows.

We observe that Aurora and BBR yield the worst TCP friendliness (10$\times$ to 60$\times$ higher throughput than Cubic). Aurora's unfriendliness results from its extremely aggressive behavior. BBR's aggressiveness aligns with results in \cite{vivace}.
Vivace yields apparent throughput disadvantages compared with Cubic, as it essentially operates like a delay-based scheme. On the other hand, \sys{} generally achieves acceptable friendliness ratios to Cubic. The reason is that \sys{} has learned an adaptive policy: it shows more tolerance to latency inflation when occupying low bandwidth, as shown in Fig.~\ref{fig:exp:hood}, which accounts for its higher throughput than delay-based schemes. However, \sys{} still avoids high latency inflation and potential packet loss, therefore it is not as aggressive as BBR and Aurora.
The experimental results demonstrate that the benefits of \sys{} do not come with aggressive bandwidth grabbing from other flows.


\subsubsection{Real-world Experiments}\label{sec:exp:perf:real-world}
\begin{figure}[t!]
    \centering
    \begin{subfigure}[b]{0.49\linewidth}
        \centering
        \includegraphics[width=\linewidth]{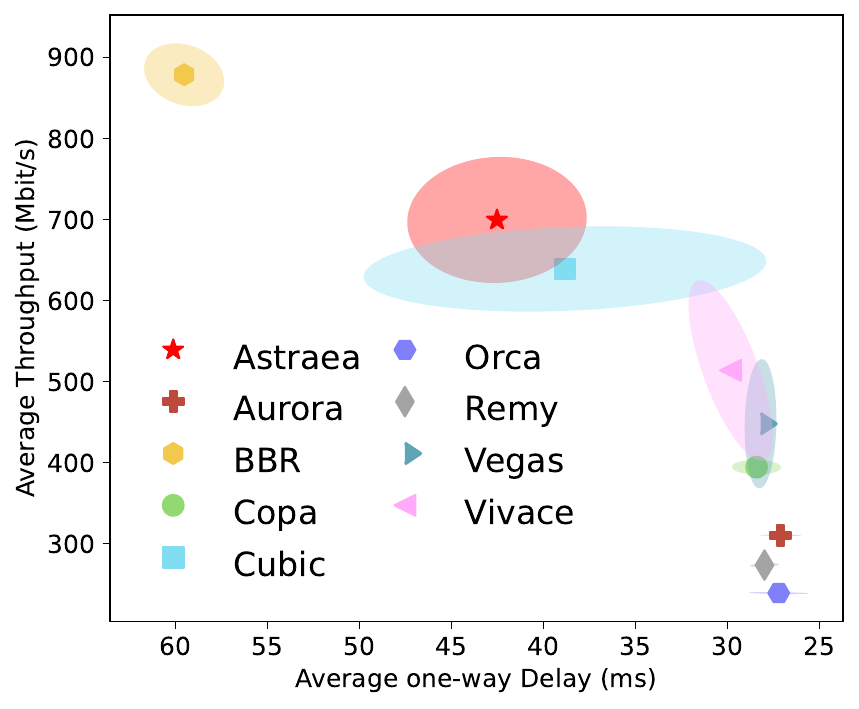}
        \caption{Intra-Continental.}
        \label{fig:exp:aws:intra}
    \end{subfigure}
    \begin{subfigure}[b]{0.49\linewidth}
        \centering
        \includegraphics[width=\linewidth]{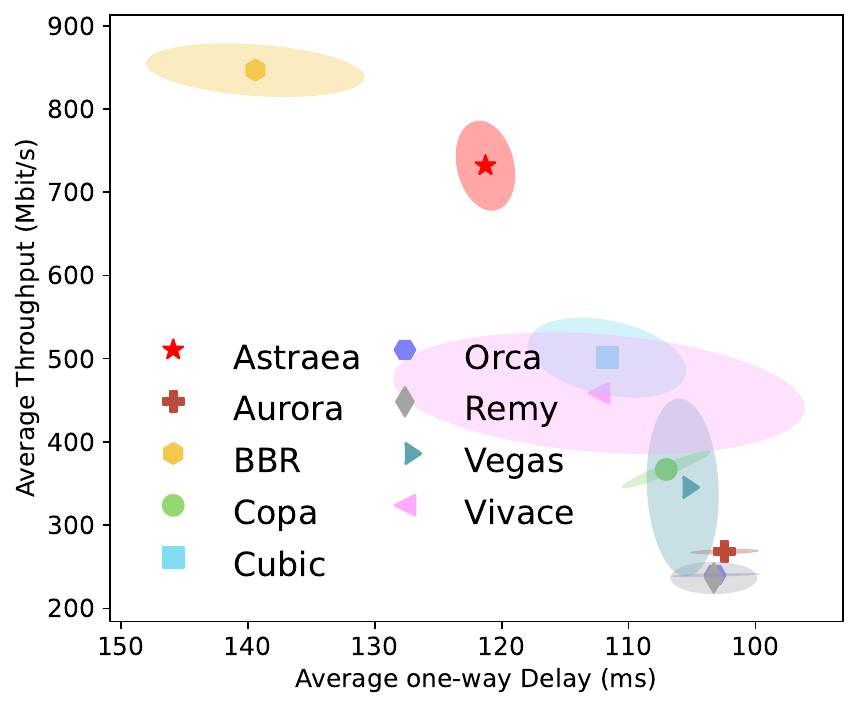}
        \caption{Inter-Continental.}
        \label{fig:exp:aws:inter}
    \end{subfigure}
    \caption{Overall average throughput vs one-way delay in real-world experiments.}
    \label{fig:exp:real}
\end{figure}

To understand how \sys performs over wide-area Internet paths with real complex traffic and sophisticated network scheduling policies, we deploy \sys with other comparisons using Pantheon~\cite{pantheon} and evaluate them on the wild Internet. We deploy senders at our residential networks and receivers at AWS nodes. We separate experiments into two categories, depending on the distance between our residential network and AWS nodes: Intra-continent and Inter-continent. We evaluate each CC scheme by running one flow for 60 seconds and repeat each trial 10 times. We summarize the overall average throughput and one-way latency in Figure~\ref{fig:exp:real}. The ellipse depicts the statistical values of total 10 trials.

%

We observe that \sys performs as the frontiers among these two kinds of experiments. For instance, in the inter-continental scenario, \sys delivers the average throughput of 731.8Mbps, which outperforms Vivace (458.5) by 1.6$\times$ and Orca (236.3) by 3.1$\times$. Among other schemes, BBR achieves the highest throughput while causing explicit latency inflation. Clean-slate machine learning-based schemes, such as Remy, Aurora and Orca do not achieve high utilization, which may due to the gap between their training environments and the wild Internet. Real-world experimental results demonstrate the generalization ability of \sys, which means, even trained under the limited range of emulated environments, \sys exhibits promising performance in the wild Internet. The high performance results from two folds. First, \sys' fast convergence speed in the single flow scenario ensures that it can quickly adapt to the changing network capacity and thus achieves high utilization. Second, \sys's reward (\S\ref{sec:design:agent}) trades a small amount of latency for higher throughput.
\subsection{Overhead}\label{sec:exp:overhead}

\begin{figure}[t!]
	\begin{subfigure}[b]{\linewidth}
        \centering
		\includegraphics[width=0.8\columnwidth]{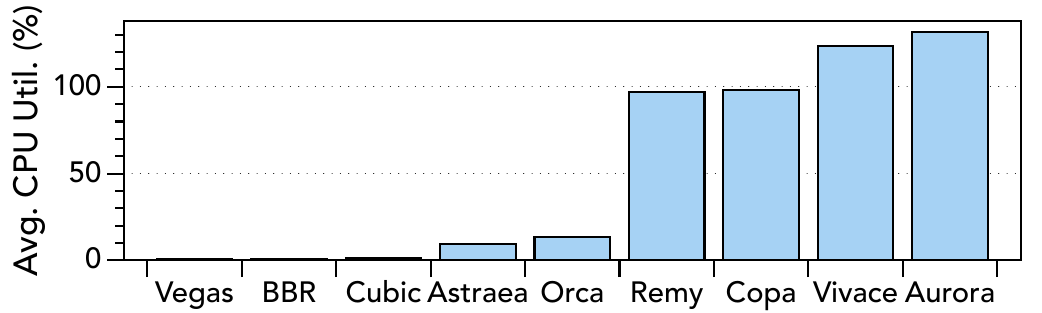}
		\caption{Average CPU utilization of each CC scheme.}
		\label{fig:exp:cpu-utilization}
    \end{subfigure}
    \begin{subfigure}[b]{\linewidth}
        \centering
        \includegraphics[width=0.8\linewidth]{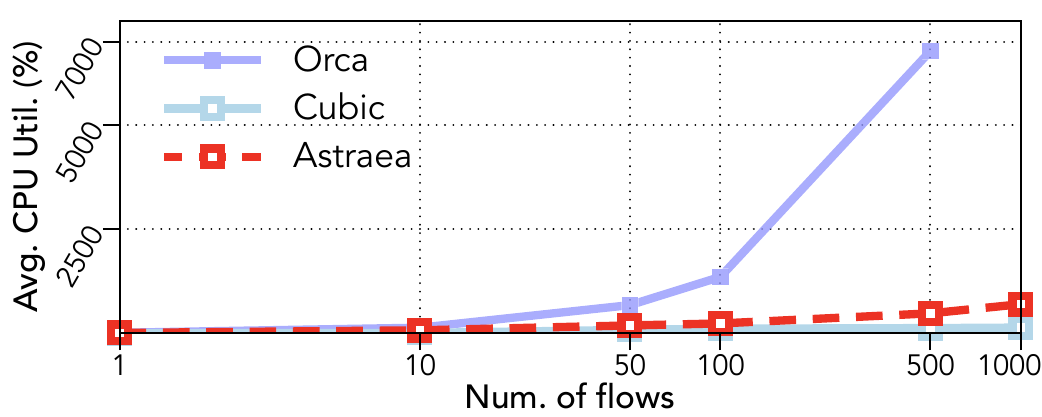}
        \caption{Scalability results of \sys{}.}
        \label{fig:exp:cpu-scale}
    \end{subfigure}
	\caption{\sys{}'s CPU overhead and scalability.}
\end{figure}

\parab{CPU utilization.}
To understand the computation overhead of \sys and demonstrate its practicability, we calculate the CPU utilization of each CC algorithm under an emulated link of 100Mbps bandwidth, 30ms RTT and 1 BDP buffer. We set the MI of \sys to be 20ms, aligning with Orca's default choice. We run each CC for 120s and report the average CPU utilization in Figure~\ref{fig:exp:cpu-utilization}. \sys achieves lower computation overhead than Orca, reducing the overhead by 30\%. \sys{}'s low overhead results from the more efficient implementation of inference service in C++. We also note that the overhead of \sys{} can be further optimized by the hierarchical design~\cite{spine} and in-kernel model execution~\cite{liteflow}, which we leave as future work.

\parab{Scalability.}
We further understand \sys{}'s scalability in Figure~\ref{fig:exp:cpu-scale}. Orca's overhead almost scales linearly with the increasing number of flows, as it requires spawning many inference server instances, which are very resource-inefficient. Our 80-core CPU machine even cannot support 1000 Orca flows. Compared to Orca, the overhead of \sys{} does not scale out linearly. This is because \sys{}'s inference service is able to serve multiple flows, as it executes the inference tasks in the batch manner, which therefore effectively mitigates the overall overhead. We note that \sys{}'s performance is not impacted by the potential response delay introduced by batch inference ($\sim$2ms), as \sys{}'s control on $cwnd$ is not sensitive to fine-grained timing interval.
This result shows that \sys{} can potentially scale out to a large number of concurrent flows.

\subsection{Interpreting \sys's Policy}\label{sec:exp:hood}
\begin{figure}[t!]
	\centering
	\includegraphics[width=0.8\columnwidth]{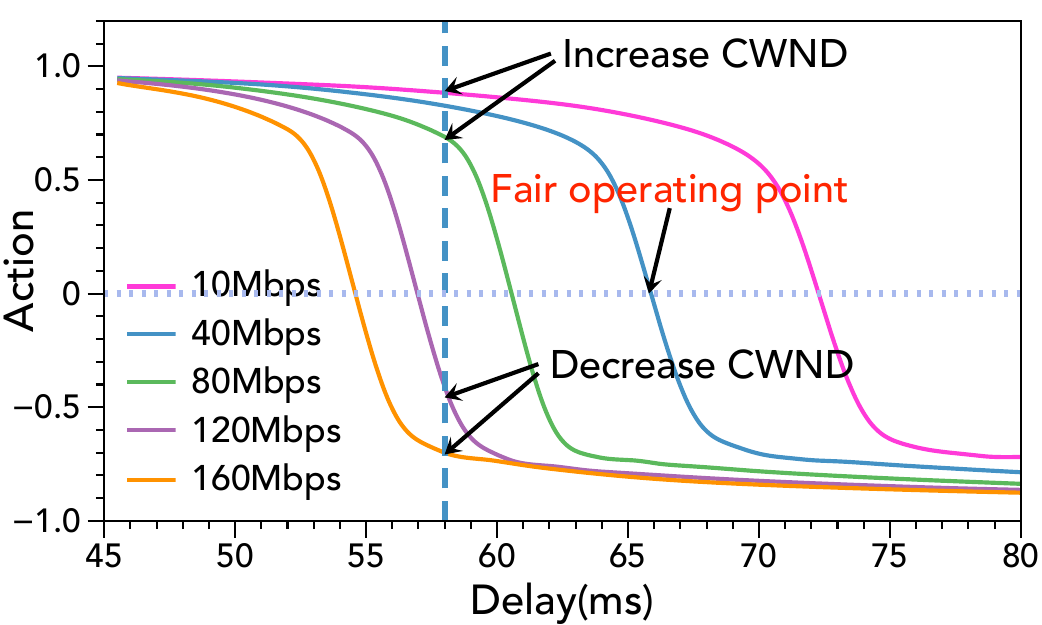}
	\caption{Details of \sys's action for competing flows.}
	\label{fig:exp:hood}
\end{figure}

In this part, we open the black box of DRL model and seek to understand how \sys achieves fairness across multiple competing flows and how it generalizes the learned policy to more diverse network conditions while preserving convergence properties by visualizing the learned state-action mappings. We fix the max-observed throughput as 200Mbps, the base RTT as 40ms, and plot the state-action mapping of various flows with varying bandwidths and delays in Figure~\ref{fig:exp:hood}. 

Our observation is that with the increasing observed delay, all \sys flows learn to lower the model output actions, transforming from $CWND$ increase to reduction. If there is only one flow in the bottleneck, it will finally reach the equilibrium point of delay where the model action is 0. Moreover, we observe that flows with different bandwidths have different equilibrium points of delay: it increases monotonically with respect to the flow bandwidth. Thus, as all the competing flows on the same bottleneck share the same queuing delay, their divergent sending rate adjustments between flows will cause the bandwidth to transfer from high-throughput flows to low-throughput flows and finally lead to a consensus on the fair operating point, where all flows equally share the link bandwidth with no more sending rate adjustment (action = 0). It can be inferred that as long as the monotonicity stays valid across all feasible bandwidths in the Internet, the fairness property of \sys under arbitrary numbers of competing flows should be guaranteed, as illustrated in Figure \ref{fig:exp:fair:large-flow}.  

\section{Related Work}\label{sec:related-work}

We have already discussed Vivace~\cite{vivace} and Aurora~\cite{aurora} in previous sections. In the following, we briefly review classical TCPs and other recently proposed schemes. 

\parab{Heuristic-based schemes:} The core of traditional TCP congestion control research is to figure out what is the congestion signal and how to respond to it. The first category of congestion signal is packet loss. TCP Tahoe, TCP Reno~\cite{reno}, TCP NewReno~\cite{newreno} are the pioneers of loss-based AIMD algorithms. Their followers spent efforts on boosting the speed of linear increasing. (\eg TCP Cubic~\cite{cubic}). Another well-known congestion signal is delay. TCP Vegas~\cite{vegas} leverages delay inflation as the signal to detect and control congestion. Latter schemes such as TCP Compound~\cite{compound} utilize a hybrid model that reacts to both delay and loss. Copa~\cite{copa} is a new delay-based scheme that aims to adjust the congestion window depending on a target rate built from network utility maximization (NUM)~\cite{num}.

\parab{Other learning-based schemes:} Orca~\cite{orca} proposes to couple classical TCP and RL for high practicability. Remy~\cite{remy} leverages offline optimizations to search the best congestion control logic under a predefined range of network conditions. Indigo~\cite{pantheon} utilizes imitation learning to train a deep neural network that encodes mapping from network observations to congestion window adjustments. While Orca's design allows it to integrate with various TCPs, the coupling may inherit the well-known issues of TCPs such as responding to non-congestive loss and TCP unfriendliness. For RL agent, its unawareness of the underlying TCP scheme's behaviors results in difficulties for adapting the learned policies in different networks, where the TCP may behave unpredictably. Besides, the learning method in Orca aims to speed up the training process by parallelly collecting each actor's local experience and optimizes the performance objectives. \sys{} differs from it through leveraging the centralized multi-agent RL to study the interactions among flows within the same network, and enabling learning and optimization on the global convergence properties while maintaining high performance.




\section{Conclusion}

We proposed \sys, a congestion control algorithm that aims to directly improve fairness, responsiveness, and stability. \sys leverages multi-agent deep reinforcement learning to build a novel paradigm for improving the convergence properties of congestion control algorithm. \sys can be naturally coupled with Linux kernel TCP and is hence readily-deployable. Extensive experimental results demonstrate that \sys shows agile responsiveness and improves fairness and stability significantly in multi-flow scenarios, while maintaining comparable performance with other schemes.



\section*{Acknowledgments}
We thank the anonymous EuroSys reviewers for their constructive feedback and suggestions. This work is supported in part by Hong Kong RGC TRS T41-603/20R, GRF 16213621, ITF ACCESS, NSFC 62062005, Key-Area Research and Development Program of Guangdong Province (2021B0101400001), and the Turing AI Computing Cloud (TACC)~\cite{TACC}. Kai Chen is the corresponding author.

\bibliographystyle{plain}
\bibliography{main}

\appendixpage
\appendix


%

\section{Training Hyperparameters and Optimizations}\label{appendix:training}

\begin{figure}[t]
	\centering
	\includegraphics[width=0.8\columnwidth]{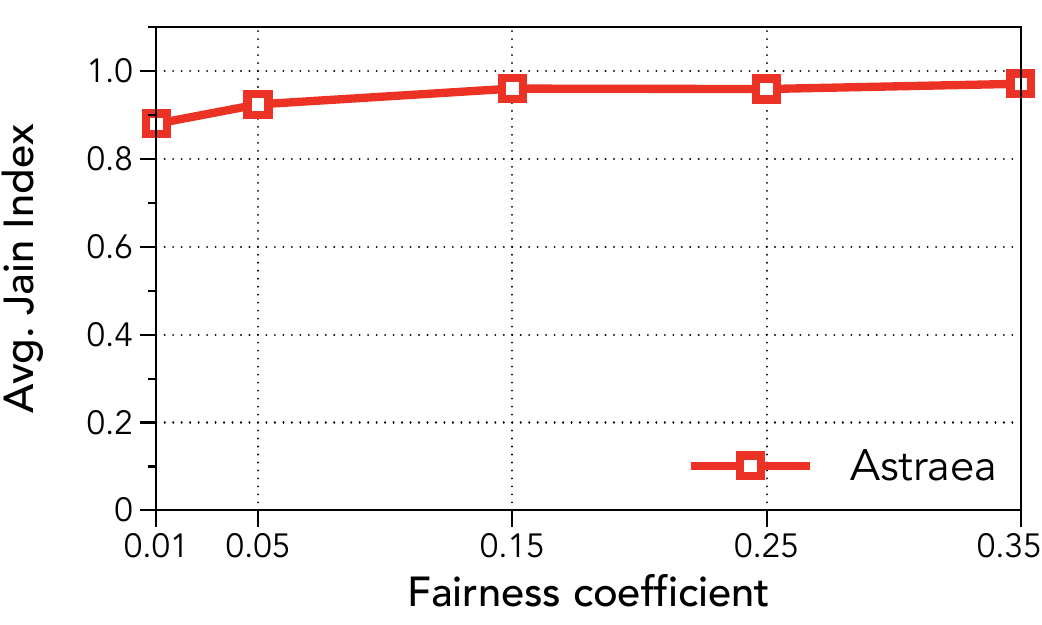}
	\caption{Fairness coefficient sensitivity}
	\label{fig:exp:fair-sensi}
\end{figure}

The detailed training hyperparameters are shown in Table \ref{table:params}.
We adopt several optimizations for \sys's multi-agent training algorithm. First, we adopt the experience replay memory technique~\cite{dqn} to smooth the learning process. The gathered tuples from agents $(g,s_i,a_i,g^\prime,s_i^\prime,r)$ are stored in a replay buffer, and the learning thread (denoted as learner) samples experiences from the buffer to calculate the gradients and update the networks. Second, to alleviate the function approximation error in actor and critic networks, we introduce optimizations from TD3~\cite{td3} including target networks, clipped double Q-Learning, delayed policy updates, and target policy smoothing regularization. We refer the reader to \cite{td3} and our implementation for the details of these optimizations. Besides, we vary the fairness coefficient and repeat the experiment in \S\ref{sec:exp:fair:simple} to evaluate the sensitivity of this parameter. Figure~\ref{fig:exp:fair-sensi} plots the average Jain index. Our observation is that \sys's fairness performance is not highly sensitive to its coefficient $c_{3}$ within the range (0.05, 0.35). 

\parab{Parallel training across flow agents.}
The multi-agent training environment in \sys naturally supports parallel training, as multiple agents act and collect statistics simutaneouly in the simulated link. 
In practice, each agent interacts with the simulated link and each other, and generate experiences in parallel. This asynchronous training framework has been widely used in reinforcement learning area and can effectively reduce the training time~\cite{a3c}. For a large-scale distributed training, we also supports multiple training environment instances, among which the same actor and critic networks are shared. Our evaluation model is trained with 4 environments instances.

\begin{table}[t!]
\centering
\begin{tabular}{c c}
\toprule
Name & Value \\
\hline
learning rate ($\alpha$) & 0.001 \\
history length ($w$) & 5 \\
gamma ($\gamma$) & 0.98 \\
batch size & 192 \\
model update interval (second) & 5 \\
model update step & 20 \\
action control coefficient ($\alpha$) & 0.025 \\
reward coefficient $c_0$ & 0.1 \\ 
reward coefficient $c_1$ & 0.02 \\
reward coefficient $c_2$ & 1 \\
reward coefficient $c_3$ & 0.02 \\
reward coefficient $c_4$ & 0.01 \\
monitoring time period (ms) & 30 \\

\bottomrule
\end{tabular}
\caption{Training parameters}
\label{table:params}
\end{table}

\section{Consistent High Performance}\label{sec:exp:emu}
\subsection{Emulated Networks with Varying Buffer Size}\label{sec:exp:emu:buffer}

\begin{figure*}[t!]
    \centering
	\begin{subfigure}[b]{0.32\linewidth}
        \centering
		\includegraphics[width=\columnwidth]{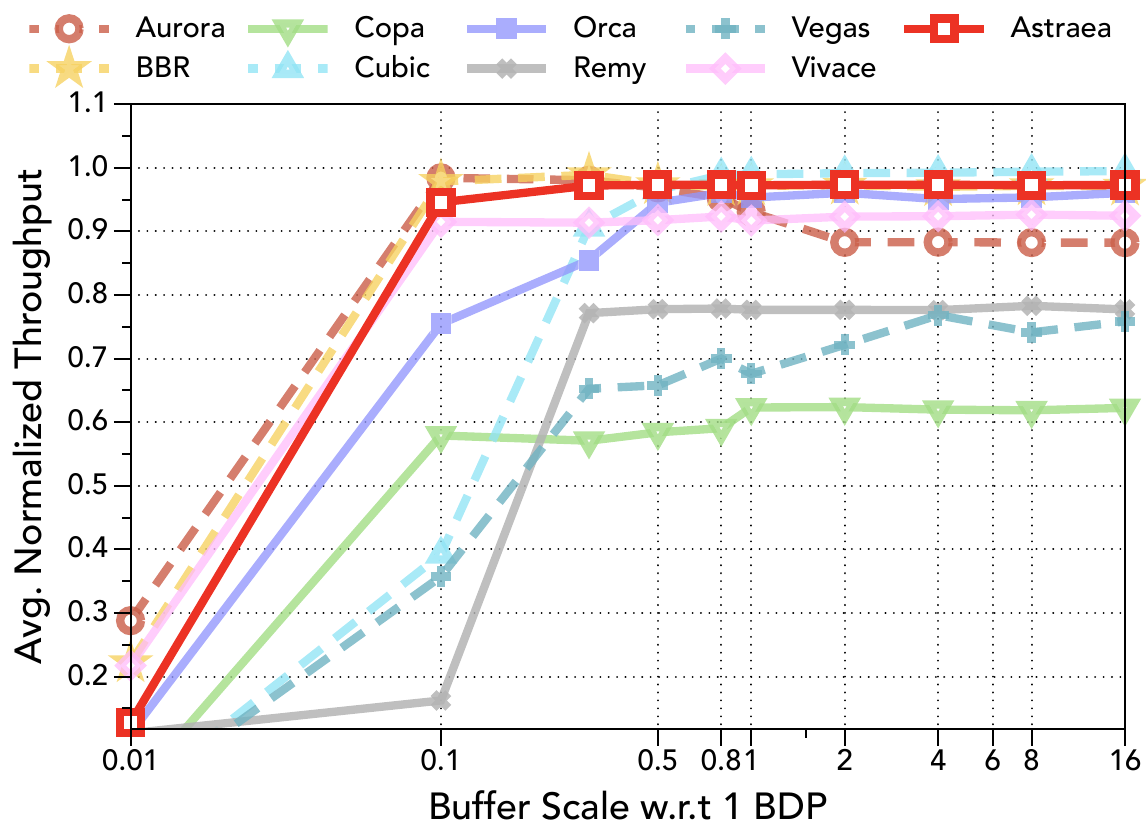}
        \caption{High Throughput.}
        \label{fig:exp:buffer:tput}
    \end{subfigure}
    \begin{subfigure}[b]{0.32\linewidth}
        \centering
        \includegraphics[width=\linewidth]{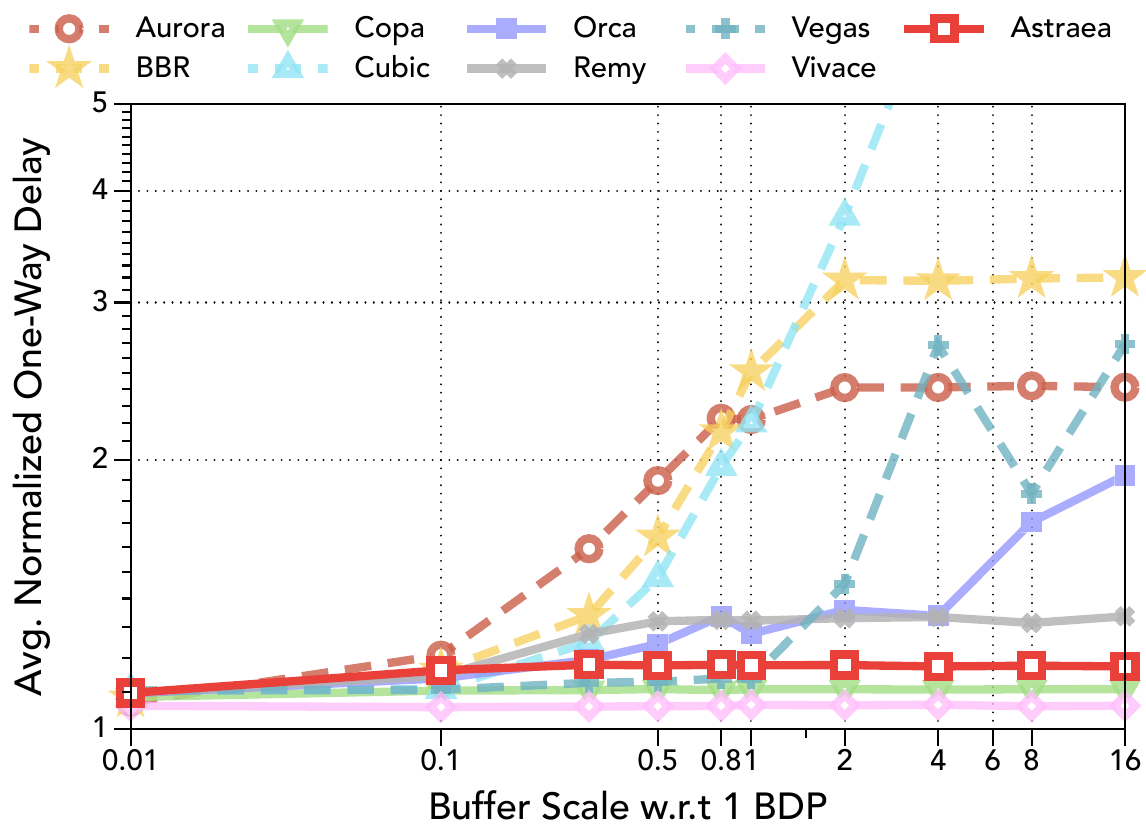}
        \caption{Mediate Latency Inflation.}
        \label{fig:exp:buffer:delay}
    \end{subfigure}
    \begin{subfigure}[b]{0.32\linewidth}
        \centering
        \includegraphics[width=\linewidth]{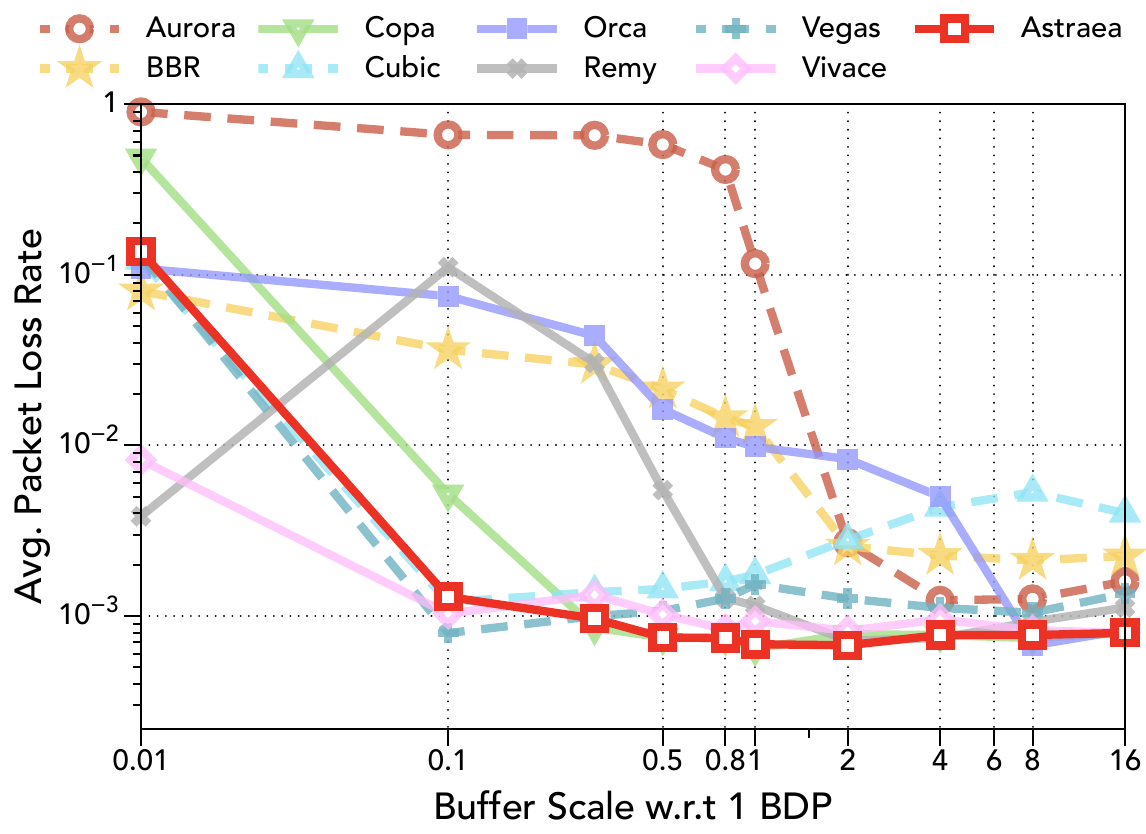}
        \caption{Low Packet Loss.}
        \label{fig:exp:buffer:loss}
    \end{subfigure}
    \caption{\sys is resilient to varying buffer size. Note that all x-axis are log-scaled and y-axis in \ref{fig:exp:buffer:delay} and \ref{fig:exp:buffer:loss} are log-scaled. Throughput and latency are normalized to link capacity and base RTT respectively.}
\end{figure*}

To understand how \sys behaves in terms of performance goals, we evaluate the performance of throughput, latency, and observed loss rate on an emulated network with 100 Mbps bandwidth, 30 ms RTT and varying buffer size. There is no random loss rate assigned. We repeat the average results of 10 trials.

On the throughput part, as depicted in Figure~\ref{fig:exp:buffer:tput}, \sys demonstrates almost full link utilization when buffer size is larger than 0.1$\times$ BDP, which is comparable with aggressive schemes such as BBR and Aurora. Orca is coupled with Cubic and thus relying on a deeper buffer (0.8$\times$ BDP) for high utilization. The delay-based schemes such as Copa and Vegas show mediate utilization as expected.

Then from the latency part in Figure~\ref{fig:exp:buffer:delay}, Vivace and delay-based schemes exhibit consistent low latency inflation as the buffer size increases. Aurora and BBR show significant latency inflation with the buffer size increasing. Orca also shows high latency when buffer size greater than 4$\times$ BDP. The reason may be that in large BDP settings, Orca's RL algorithm is dominated by the underlying TCP Cubic logic, which is well-known for the buffer-filling behavior. \sys achieves slightly higher latency compared with Vivace as the cost of high throughput.

Finally, we compare the average packet loss in Figure~\ref{fig:exp:buffer:loss}. Orca does not perform well in shallow buffer settings given that it even produces a higher loss rate than Cubic when buffer size is less than 4$\times$ BDP, since it does not optimize for responsiveness in shallow buffers. \sys only requires 0.1$\times$ BDP to perform a near-lossless data transfer. Specifically, when buffer size is larger than 0.1$\times$ BDP, \sys consistently delivers one of the lowest packet loss rates across all evaluated schemes.

\subsection{Unreliable Networks with Large RTT}\label{sec:exp:emu:satellite}
\begin{figure}[t!]
	\centering
	\includegraphics[width=0.8\columnwidth]{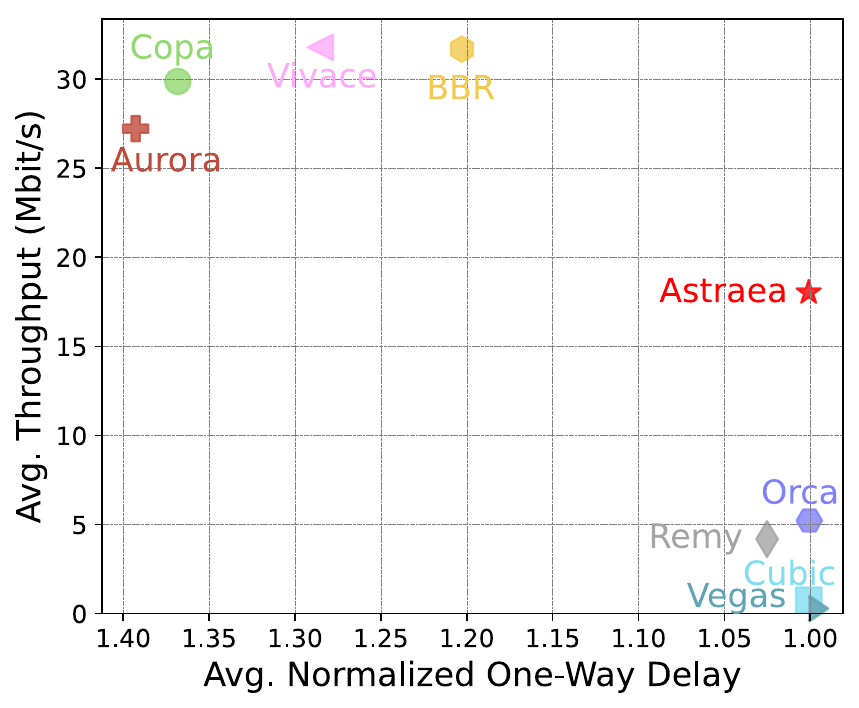}
	\caption{Throughput vs normalized delay plot for the satellite link. Note that the loss-insensitive schemes tend to do well on throughput, while delay-based schemes do well in normalized delay.}
	\label{fig:exp:satellite}
\end{figure}
We evaluate \sys on an emulated satellite link using the setting suggested in Vivace Paper~\cite{vivace}. Satellite link is challenging because of the long RTT, corresponding high BDP and random loss. We setup the link with 42 Mbps bandwidth, 800ms RTT, 1 BDP buffer size with 0.74\% stochastic loss rate, where each sender runs for 100 seconds. We repeat each test with each CC scheme 10 times and report the average throughput and normalized one-way latency in Figure~\ref{fig:exp:satellite}. We observe that Cubic and Vegas respond to loss thus leading to low throughput. Orca is impacted by underlying TCP Cubic and demonstrates the similar performance. Long RTT exhibits impact on BBR's probing process and therefore make its rate oscillated, which accounts for the high utilization with high latency. Both Vivace, Copa, and Aurora do not respond to packet loss and hence achieve high throughput. \sys is trained to be loss-resilient and employs similar logic to respond to loss as BBR, which demonstrates the moderate throughput and low latency.

\begin{figure}[t!]
	\centering
	\includegraphics[width=0.8\columnwidth]{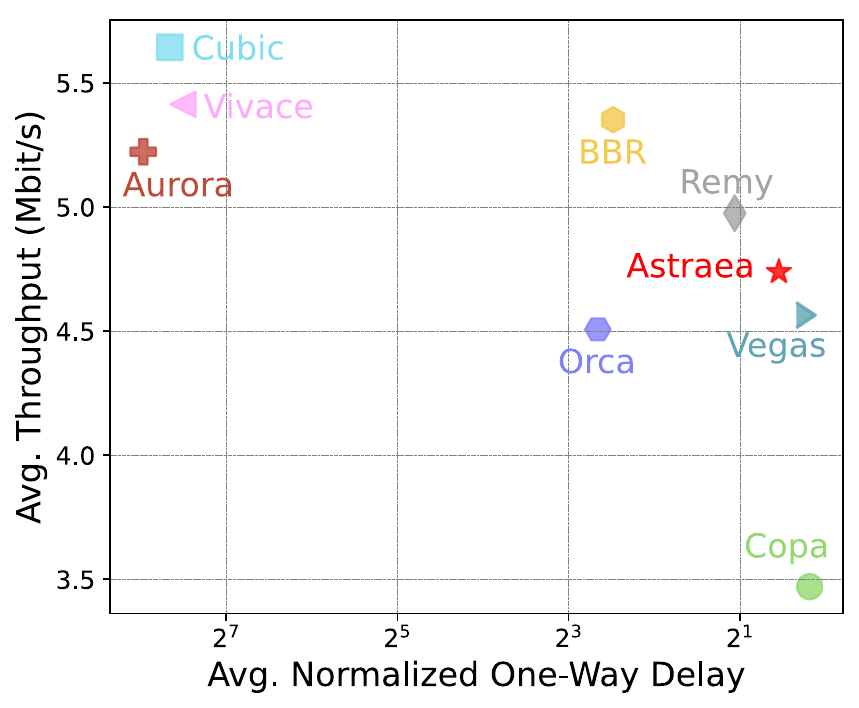}
	\caption{Throughput vs normalized delay plot for the cellular link using the Verizon LTE trace. Latency is normalized to the base RTT. Note that x-axis is log-scaled. 
	}
	\label{fig:exp:cellular}
\end{figure}

\subsection{Performance in Rapidly Changing Networks}\label{sec:exp:resp:changing}
Figure~\ref{fig:exp:cellular} reports the average statistical results  of each CC under cellular networks. In general, \sys maintains high throughput and low latency inflation. Other learning-based schemes, such as Aurora and Vivace, can achieve high throughput at the cost of obviously high latency. The delay-based schemes, Copa and Vegas, demonstrate low latency but sacrificing utilization. 

\begin{figure}[t!]
	\centering
	\includegraphics[width=0.8\columnwidth]{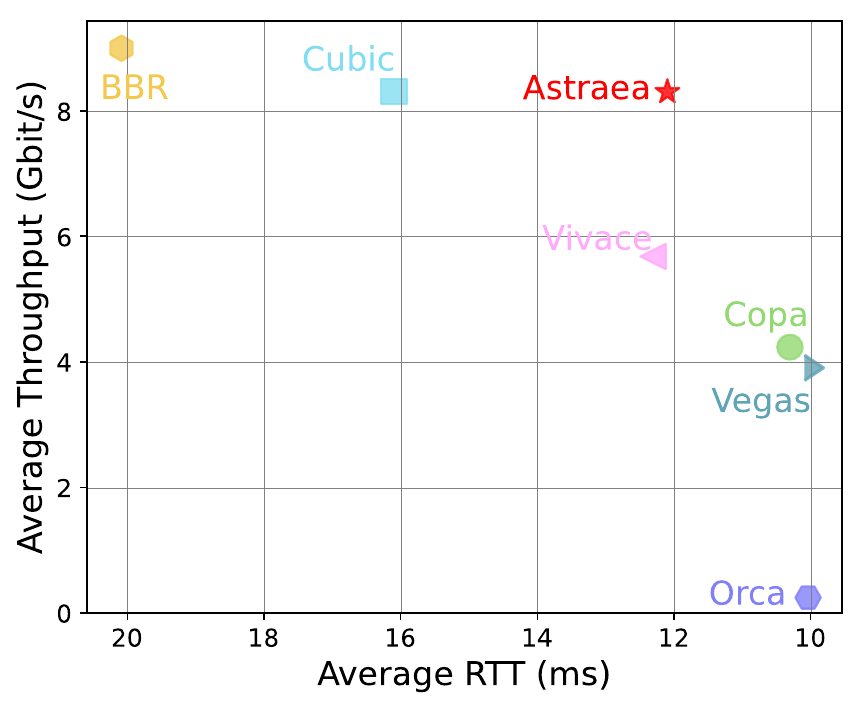}
	\caption{Throughput vs RTT plot for 10Gbps WAN network}
	\label{fig:exp:dcn}
\end{figure}
\subsection{Performance in High-speed WAN Networks}\label{sec:exp:high-speed}
We also evaluate \sys{} in an emulated high-speed WAN network with 10Gbps bandwidth and 10ms base RTT using Linux TC~\cite{tc}. Figure~\ref{fig:exp:dcn} reports the average results of each CC scheme. We find that \sys{} delivers higher throughput than Orca and Vivace. \sys{}'s fast convergence to link bandwidth ensures its high utilization. In addition, \sys{}'s low latency results from the penalty of high latency inflation in its reward function.

\end{document}